# Light propagation in a horizontally homogeneous Lambertian foliage: Analytically solvable models

Antoine Royer

*Département de Génie Physique, Ecole Polytechnique, Montréal, Québec H3C 3A7, Canada*

Sophie Royer

*Institut für Botanik, Universität Innsbruck, Austria*

Various numerical methods exist for obtaining the radiances inside a canopy of leaves above a partly reflecting ground. In view of testing the accuracy of these diverse methods, it is desirable to have at one's disposal non-trivial models possessing analytical solutions, against which to compare numerical results. Such models are obtained in the present paper, for the case of a horizontally homogeneous foliage of Lambertian leaves, modeled as a turbid medium. Our treatment is more general than usual in that we allow the top and under sides of leaves to have different optical coefficients. Besides being more realistic, this enables artificial situations, such as extreme light trapping, testing the limits of the various numerical methods.

## 1. Introduction

The problem of light propagation in foliage is of prime importance in various applied fields of research, such as remote sensing, ecology, and agriculture [1-4]. The light climate inside a canopy, as well as the light reflected to space, are usually obtained by numerically integrating a *light transport equation* (LTE) [1-4].

Our aim is to eventually compute rates of photosynthesis inside canopies. These rates interact with the leaf temperatures, and both depend *non-linearly* on the rates of absorption of radiation by leaves. The latter in turn depend on the ambient radiances, hence on thermal emissions by leaves, which themselves depend non-linearly (again!) on leaf temperatures. Thus it appears that the only feasible way to determine leaf temperatures is by iteration. Whence the necessity of repeatedly integrating the LTE, which it is therefore imperative to do in minimum time. The most widely used numerical integration method for doing so is an iterative procedure [1-3], which we call 'iterative integration' (this is also often called 'relaxation method' [1]). This iterative integration (which comes on top of the iterations over temperatures) may entail some loss of accuracy, and be time-consuming in certain situations.

In a preceding paper [5], we proposed a new method for integrating transport equations, by combining transfer matrices, transmission-reflection matrices, and Green's matrices (TTRG).

This method turns out to be both faster and more accurate than iterative integration [1-3]. In order to test the accuracy and limits of our method, as well as its speed, and compare it with other methods, it is desirable to have at one's disposal non-trivial models possessing analytic solutions, against which to compare numerical results. The main purpose of the present paper is to provide such models (which are also of intrinsic interest). These will be used in subsequent papers to test the reliability of our method. Another purpose of the paper is to set up the basic theory for subsequent papers presenting realistic calculations.

As is usual, the canopy is modeled as a turbid medium of flat leaf elements of zero thickness and infinitesimal areas [1-4]. However, our treatment is more general than usual in that we allow the top and bottom sides of leaves to have different optical properties. This is more realistic. Also, it allows us to create artificial situations, such as extreme light trapping, to test the limits of numerical methods. It will turn out that in cases of extreme light trapping, iterative integration can become impractical, while our method remains as efficient

We restrict ourselves to a horizontally homogeneous canopy. Also, we do not distinguish between photons of different frequencies or polarizations (although in papers dealing with rates of photosynthesis, we shall, as is usual [1-4], treat separately visible, near infrared, and thermal photons). The theory is set up for a canopy comprising a single species of leaf, the generalization to several species being straightforward.

The paper has three parts. Part I defines the basic quantities of the theory, and derives the light transport equation, and variants of it. Part II assumes purely Lambertian scattering and emission (specular reflection will be incorporated in later papers). Part III formulates analytically solvable models. Some technical details are relegated to two Appendices.

***Notations and conventions***: In order to make equations less cluttered and more readable, we will often use a hat over scalars $c$ to signify *absolute value*: Thus, $\hat{c} \equiv |c| \geq 0$. Using the step function $\Theta(x)$, we also denote

$$|x|_{\pm} \equiv \hat{x}\,\Theta(\pm x), \qquad \hat{x} \equiv |x|, \qquad \Theta(x>0)=1, \quad \Theta(x<0)=0 \tag{1}$$

We let the positive $z$ direction point *downwards* (i.e., towards the ground surface). The unit vector in direction $z$ is denoted $\mathbf{u}_z \equiv (0,0,1)$. Unit vectors in direction $\Omega = (\mu, \varphi)$, where

$\mu \equiv \cos\theta$, are denoted:

$$\boldsymbol{\Omega} = (\Omega_x, \Omega_y, \Omega_z) = (\sin\theta\cos\varphi, \sin\theta\sin\varphi, \mu), \qquad \mu = \Omega_z = \cos\theta$$

$$d\Omega = \sin\theta d\theta d\varphi = d\mu d\varphi, \qquad \int d\Omega f(\Omega) = \int_{-\pi}^{\pi} d\varphi \int_{-1}^{1} d\mu\, f(\mu, \varphi) \tag{2}$$

The scalar product of two unit vectors $\boldsymbol{\Omega} = (\mu, \varphi)$ and $\boldsymbol{\Omega}' = (\mu', \varphi')$ is

$$\boldsymbol{\Omega} \cdot \boldsymbol{\Omega}' = a + b\cos(\varphi - \varphi'), \qquad a \equiv \mu\mu', \qquad b \equiv \sin\theta\sin\theta' \geq 0 \tag{3}$$

We will need the integrals

$$\int d\Omega |\boldsymbol{\Omega} \cdot \boldsymbol{\Omega}'|_{\pm} = \int d\Omega |\mu|_{\pm} = \int_{-\pi}^{\pi} d\varphi \int_{0}^{1} \mu d\mu = \pi$$

$$\int d\Omega |\boldsymbol{\Omega} \cdot \boldsymbol{\Omega}'| = 2\pi, \qquad \tfrac{1}{2\pi} \int_{-\pi}^{\pi} d\varphi\, \boldsymbol{\Omega} \cdot \boldsymbol{\Omega}' = a = \mu\mu' \tag{4}$$

The angular $\delta$ function $\delta(\Omega - \Omega_0) = \delta(\mu - \mu_0)\delta(\varphi - \varphi_0)$ is such that

$$\int_{\Delta\Omega} d\Omega f(\Omega)\delta(\Omega - \Omega_0) = \begin{cases} f(\Omega_0) & \text{if } \Omega_0 \in \Delta\Omega \\ 0 & \text{if } \Omega_0 \notin \Delta\Omega \end{cases} \tag{5}$$

where $\Delta\Omega$ is some solid angle.

We define functions (with $a, b$ given by (3), and $\hat{\mu} \equiv |\mu|$, etc.):

$$g(\mu, \mu') \equiv \tfrac{1}{2\pi} \int_{-\pi}^{\pi} d\varphi |\boldsymbol{\Omega} \cdot \boldsymbol{\Omega}'| = \begin{cases} \hat{\mu}\hat{\mu}' & \text{if } \hat{a} \geq \hat{b} \\ \tfrac{2}{\pi}\zeta(-a/b) - a & \text{if } \hat{a} \leq \hat{b} \end{cases} \tag{6}$$

$$\zeta(\mu) \equiv \sqrt{1-\mu^2} - \mu\cos^{-1}\mu = -\int d\mu \cos^{-1}\mu \tag{7}$$

(see Appendix B.1). We also define, letting $u$ stand for $+$ or $-$:

$$g_{u_i, u_f}(\mu_i, \mu_f, \mu_L, \varphi_{if}) \equiv \tfrac{1}{2\pi} \int_{-\pi}^{\pi} d\varphi_L |\boldsymbol{\Omega}_i \cdot \boldsymbol{\Omega}|_{u_i} |\boldsymbol{\Omega}_f \cdot \boldsymbol{\Omega}|_{u_f}, \qquad \varphi_{if} \equiv \varphi_i - \varphi_f \tag{8}$$

for which a complicated analytic form can be obtained (see Appendix B.2).

We write $\boldsymbol{\Omega} \in D$ if $\mu > 0$ ($\boldsymbol{\Omega}$ points downwards), and $\boldsymbol{\Omega} \in U$ if $\mu < 0$ (upwards). A function $f(\Omega) = f(\mu)$ independent of $\varphi$ is called azymuthally symmetric, or $\varphi$-symmetric for short. Infinitesimal lengths, areas, and volumes are denoted $d\ell$, $da$, and $d^3\mathbf{r}$, respectively.

**Part I  General theory**

This part concerns the general aspects of the theory. Section 2 defines the basic quantities of the theory. Section 3 derives the light transport equation (LTE). Section 4 discusses extinction coefficients, and writes them as a sum of two terms, due respectively to interception by leaf tops, and by leaf bottoms. Section 5 restricts the theory to horizontally homogeneous situations. Section 6 specifies the boundary conditions at the ground surface, and states global energy balance. Finally, section 7 treats first-scattered-sunlight as an 'emission'.

**2  Basic quantities**

This section introduces the basic quantities of the theory. Some definitions are displayed for easy reference.

***Photon flux (radiance)***: We refer to a photon travelling in direction $\mathbf{\Omega}$, or within $d\Omega$ about $\mathbf{\Omega}$, as a 'photon $\mathbf{\Omega}$', or a 'photon $d\Omega$'. The photon flux $I(\mathbf{\Omega},\mathbf{r})$ is defined by:

$$I(\mathbf{\Omega},\mathbf{r})d\Omega da_\Omega = \begin{cases} \text{number of photons } d\Omega \text{ crossing per second an infinitesimal} \\ \text{area } da_\Omega \text{ perpendicular to } \mathbf{\Omega} \text{ at position } \mathbf{r} = (x,y,z). \end{cases} \quad (9)$$

We will also refer to $I(\mathbf{\Omega},\mathbf{r})$ as the *radiance* (though more properly, radiance is the *energy* flux). The set of radiances $I(\mathbf{\Omega},\mathbf{r})$ for all $\mathbf{\Omega}$ is called the *light climate* at $\mathbf{r}$.

***Leaf orientations and leaf area densities***: The canopy is modeled as a turbid medium of flat leaf elements of infinitesimal areas. *Horizontal* leaf elements are assumed azymuthally isotropic (i.e., $\varphi$-symmetric), so that it suffices to specify the orientation of an *inclined* leaf element by the normal $\mathbf{\Omega}_L$ to its *bottom* surface. In the case of a horizontal leaf, $\mathbf{\Omega}_L = \mathbf{u}_z$, where $\mathbf{u}_z$ is the unit vector in direction $z$ (downwards). We refer to a leaf with normal $\mathbf{\Omega}_L$ as a 'leaf $\mathbf{\Omega}_L$'.

The canopy is characterized by a leaf area density $\eta(\mathbf{\Omega}_L,\mathbf{r})$, defined by:

$$\eta(\mathbf{\Omega}_L,\mathbf{r})d\Omega_L d^3\mathbf{r} = \text{total leaf area oriented within } d\Omega_L \text{ inside } d^3\mathbf{r} \text{ at } \mathbf{r}. \quad (10)$$

We also define $\varphi_L$ - integrated, and *total*, leaf area densities at $\mathbf{r}$:

$$\begin{aligned} \tilde{\eta}(\mu_L,\mathbf{r}) &\equiv \int_{-\pi}^{\pi} d\varphi_L \eta(\Omega_L,\mathbf{r}) \\ \eta(\mathbf{r}) &\equiv \int d\Omega_L \eta(\Omega_L,\mathbf{r}) = \int_{-1}^{1} d\mu_L \tilde{\eta}(\mu_L,\mathbf{r}) \end{aligned} \quad (11)$$

The *distribution* $\lambda(\Omega_L, \mathbf{r})$ of leaf orientations $\Omega_L$, at position $\mathbf{r}$, and the distribution $\tilde{\lambda}(\mu_L, \mathbf{r})$ of leaf inclinations $\mu_L$, are given by

$$\lambda(\Omega_L, \mathbf{r}) \equiv \eta(\Omega_L, \mathbf{r})/\eta(\mathbf{r})$$
$$\tilde{\lambda}(\mu_L, \mathbf{r}) \equiv \int_{-\pi}^{\pi} d\varphi_L \, \lambda(\Omega_L, \mathbf{r}) = \tilde{\eta}(\mu_L, \mathbf{r})/\eta(\mathbf{r}) \tag{12}$$

They satisfy the normalizations

$$\int d\Omega_L \, \lambda(\Omega_L, \mathbf{r}) = \int_{-1}^{1} d\mu_L \, \tilde{\lambda}(\mu_L, \mathbf{r}) = 1 \tag{13}$$

Thus, the average of a function $f(\Omega_L)$ over $\Omega_L$, and that of $f(\mu_L)$ over $\Omega_L$ or $\mu_L$, are:

$$\langle f(\Omega_L) \rangle_{\mathbf{r}} \equiv \int d\Omega_L \, \lambda(\mu_L, \mathbf{r}) f(\Omega_L), \qquad \langle f(\mu_L) \rangle_{\mathbf{r}} \equiv \int_{-1}^{1} d\mu_L \, \tilde{\lambda}(\mu_L, \mathbf{r}) f(\mu_L) \tag{14}$$

***Leaf optical coefficients***: A photon hitting a leaf is either scattered or absorbed. Scattering and absorption coefficients for a *horizontal* leaf, $\sigma(\Omega_i \to \Omega_f)$ and $\alpha(\Omega_i)$, are defined by:

$$\sigma(\Omega_i \to \Omega_f) d\Omega_f = \text{probability for a photon } \Omega_i \text{ to get scattered into } d\Omega_f. \tag{15}$$

$$\alpha(\Omega_i) = 1 - \int d\Omega_f \, \sigma(\Omega_i \to \Omega_f) = \text{probability to get absorbed.} \tag{16}$$

Moreover, emission coefficients $\varepsilon(\Omega)$ are defined by:

$$\varepsilon(\Omega_f) d\Omega_f = \begin{cases} \text{number of photons emitted per second} \\ \text{into } d\Omega_f \text{ by a unit area of leaf.} \end{cases} \tag{17}$$

We will often think of scattering as 'absorption' followed by 'emission'.

In the case of an inclined leaf with normal $\mathbf{\Omega}_L$, these optical coefficients are denoted $\sigma_{\Omega_L}(\Omega_i \to \Omega_f)$, $\alpha_{\Omega_L}(\Omega_i)$ and $\varepsilon_{\Omega_L}(\Omega_f, \mathbf{r})$, where we let emission depend on $\mathbf{r}$, since the temperature of a leaf depends in general on its position inside the canopy.

## 3 Canopy optical coefficients and light transport equation (LTE)

In this section, we calculate local optical coefficients for the canopy, and then derive the light transport equation (LTE), which tells us how the photon flux $I(\Omega, \mathbf{r})$ varies in direction $\Omega$, due to scattering, absorption and emission by leaves.

***Canopy optical coefficients***: Let a photon $\Omega_i$ travel an *infinitesimal* distance $\mathbf{\Omega}_i \, d\ell$ inside the canopy. Denote its probability to hit a leaf by $\Gamma(\Omega_i, \mathbf{r}) d\ell$, and that to get scattered into $d\Omega_f$ by $S(\Omega_i \to \Omega_f, \mathbf{r}) d\Omega_f d\ell$. Since $\mathbf{\Omega}_i d\ell$ pierces at most one leaf element, and since an infinitesimal area $da_L$ of leaf $\Omega_L$ is 'seen' as $|\mathbf{\Omega}_i \cdot \mathbf{\Omega}_L| da_L$ by photons $\Omega_i$, we have, using (10) and (15):

$$\Gamma(\Omega_i, \mathbf{r}) = \int d\Omega_L \eta(\Omega_L, \mathbf{r}) |\mathbf{\Omega}_i \cdot \mathbf{\Omega}_L|$$
$$S(\Omega_i \to \Omega_f, \mathbf{r}) = \int d\Omega_L \eta(\Omega_L, \mathbf{r}) |\mathbf{\Omega}_i \cdot \mathbf{\Omega}_L| \sigma_{\Omega_L}(\Omega_i \to \Omega_f) \quad (18)$$

The probability of absorption per unit distance is the difference

$$A(\Omega_i, \mathbf{r}) = \Gamma(\Omega_i, \mathbf{r}) - \int d\Omega_f S(\Omega_i \to \Omega_f, \mathbf{r})$$
$$= \int d\Omega_L \eta(\Omega_L, \mathbf{r}) |\mathbf{\Omega}_i \cdot \mathbf{\Omega}_L| \alpha_{\Omega_L}(\Omega_i) \quad (19)$$

Letting also $\mathcal{E}(\Omega, \mathbf{r}) d\Omega d^3\mathbf{r}$ be the number of photons emitted per second into $d\Omega$ by an infinitesimal volume $d^3\mathbf{r}$ of canopy, we have:[1]

$$\mathcal{E}(\Omega, \mathbf{r}) = \int d\Omega_L \eta(\Omega_L, \mathbf{r}) \varepsilon_{\Omega_L}(\Omega, \mathbf{r}) \quad (20)$$

Thus, $\mathcal{E}(\Omega, \mathbf{r}) \mathbf{\Omega} d\ell$ is the flux of photons $\Omega$ emitted by a thickness $\mathbf{\Omega} d\ell$ of canopy.

***Scattered light as an 'emission'***: In view of (9), a number $dN_i = I(\Omega_i, \mathbf{r}) d\Omega_i da_i$ of photons $d\Omega_i$ crosses per second an area $da_i$ perpendicular to $\mathbf{\Omega}_i$. Over a distance $\mathbf{\Omega}_i d\ell$ each photon has a probability $p_{i \to f} = S(\Omega_i \to \Omega_f, \mathbf{r}) d\Omega_f d\ell$ of getting scattered into $d\Omega_f$. Hence the volume $\mathbf{\Omega}_i d\ell da_i$ scatters $dN_f = dN_i p_{i \to f}$ photons $d\Omega_i$ into $d\Omega_f$, per second. Thus, scattered photons may be viewed as due to an 'emission' per unit volume

$$\mathcal{E}^{scatt}(\Omega_f, \mathbf{r}) = \int d\Omega_i I(\Omega_i, \mathbf{r}) S(\Omega_i \to \Omega_f, \mathbf{r}) \quad (21)$$

***The light transport equation***: Consider a light beam of radiance $I(\Omega_f, \mathbf{r})$ in direction $\Omega_f$. Over an infinitesimal distance $\mathbf{\Omega}_f d\ell$, the beam looses a number $\Gamma(\Omega_f, \mathbf{r}) I(\Omega_f, \mathbf{r}) d\ell$ of photons due to interception by leaves, but gains a number $\mathcal{E}^{total}(\Omega_f, \mathbf{r}) d\ell$ of photons due to the total 'emission' $\mathcal{E}^{total} = \mathcal{E} + \mathcal{E}^{scatt}$. Whence the light transport equation (LTE)

---

[1] Note that $\mathcal{E}$ in the present paper represents other objects than in Eqs.(33) *et seq.* of Ref.[5].

$$\mathbf{\Omega}_f \cdot \nabla I(\mathbf{\Omega}_f,\mathbf{r}) = -\Gamma(\mathbf{\Omega}_f,\mathbf{r})I(\mathbf{\Omega}_f,\mathbf{r}) + \int d\Omega_i \, I(\mathbf{\Omega}_i,\mathbf{r})S(\mathbf{\Omega}_i \to \mathbf{\Omega}_f,\mathbf{r}) + \mathcal{E}(\mathbf{\Omega}_f,\mathbf{r}) \quad (22)$$

*Continuity equation*: Writing $\mathbf{\Omega}\cdot\nabla I(\mathbf{\Omega},\mathbf{r}) = \nabla\cdot\mathbf{i}(\mathbf{\Omega},\mathbf{r})$ where $\mathbf{i}(\mathbf{\Omega},\mathbf{r}) \equiv \mathbf{\Omega} I(\mathbf{\Omega},\mathbf{r})$ is the photon flux *vector* in direction $\mathbf{\Omega}$, and integrating (22) over $\mathbf{\Omega}_f$, using (19), we get a continuity equation for the *net* current $\mathbf{i}(\mathbf{r})$:

$$\nabla\cdot\mathbf{i}(\mathbf{r}) = \mathcal{E}(\mathbf{r}) - \mathcal{A}(\mathbf{r}), \qquad \mathbf{i}(\mathbf{r}) \equiv \int d\Omega\, \mathbf{\Omega} I(\mathbf{\Omega},\mathbf{r}) \quad (23)$$

where $\mathcal{E}(\mathbf{r})$ and $\mathcal{A}(\mathbf{r})$ are the total emission and absorption rates per unit volume:

$$\mathcal{E}(\mathbf{r}) \equiv \int d\Omega\, \mathcal{E}(\mathbf{\Omega},\mathbf{r}), \qquad \mathcal{A}(\mathbf{r}) \equiv \int d\Omega\, I(\mathbf{\Omega},\mathbf{r})A(\mathbf{\Omega},\mathbf{r}) \quad (24)$$

since the rate of absorption of photons $\mathbf{\Omega}$ is $I(\mathbf{\Omega},\mathbf{r})A(\mathbf{\Omega},\mathbf{r})$.

## 4 Extinction and attenuation coefficients

The functions $\Gamma(\mathbf{\Omega},\mathbf{r}) \geq 0$ defined in (18) are called *extinction coefficients*, because they determine how a light beam gets *attenuated* due to interception by leaves. We note the bound

$$\Gamma(\mathbf{\Omega},\mathbf{r}) \leq \int d\Omega_L\, \eta(\mathbf{\Omega}_L,\mathbf{r}) = \eta(\mathbf{r}) \quad (25)$$

since $|\mathbf{\Omega}\cdot\mathbf{\Omega}_L| \leq 1$ (equality corresponds to all leaves perpendicular to $\mathbf{\Omega}$).

*Attenuation coefficients*: Since the probability for a photon $\mathbf{\Omega}$ to get intercepted by a leaf over a distance $\mathbf{\Omega} d\ell$ is $\Gamma(\mathbf{\Omega},\mathbf{r})d\ell$, the probability $\xi_\Omega(\mathbf{r},\mathbf{r}+\mathbf{\Omega}\ell)$ for the photon to go unimpeded from $\mathbf{r}$ to $\mathbf{r}+\mathbf{\Omega}\ell$ satisfies $d\xi_\Omega/d\ell = -\Gamma\xi_\Omega$. Whence the attenuation coefficient

$$\xi_\Omega(\mathbf{r},\mathbf{r}+\mathbf{\Omega}\ell) = e^{-\Lambda_\Omega(\mathbf{r},\mathbf{r}+\mathbf{\Omega}\ell)} \quad (26)$$

where

$$\Lambda_\Omega(\mathbf{r},\mathbf{r}+\mathbf{\Omega}\ell) \equiv \int_0^\ell d\ell'\, \Gamma(\mathbf{\Omega},\mathbf{r}+\mathbf{\Omega}\ell') \quad (27)$$

*The coefficients* $\Gamma_\pm$: Because we allow the top and bottom sides of leaves to have different optical properties, it is useful to define separate probabilities of interception $\Gamma_+$ by leaf tops, and $\Gamma_-$ by leaf bottoms (see notation (1)):

$$\begin{aligned}\Gamma_\pm(\mathbf{\Omega},\mathbf{r}) &\equiv \int d\Omega_L\, \eta(\mathbf{\Omega}_L,\mathbf{r})|\mathbf{\Omega}\cdot\mathbf{\Omega}_L|_\pm = \tfrac{1}{2}\Gamma \pm \tfrac{1}{2}\Delta\Gamma, \qquad \Gamma = \Gamma_+ + \Gamma_- \\ \Delta\Gamma(\mathbf{\Omega},\mathbf{r}) &\equiv \Gamma_+ - \Gamma_- = \int d\Omega_L\, \eta(\mathbf{\Omega}_L,\mathbf{r})\,\mathbf{\Omega}\cdot\mathbf{\Omega}_L\end{aligned} \quad (28)$$

where we used $|x|_+ - |x|_- = x$. Since $|\boldsymbol{\Omega}\cdot\boldsymbol{\Omega}_L|_- = |(-\boldsymbol{\Omega})\cdot\boldsymbol{\Omega}_L|_+ = |\boldsymbol{\Omega}\cdot(-\boldsymbol{\Omega}_L)|_+$, we have

$$\Gamma_-(\boldsymbol{\Omega},\mathbf{r}) = \Gamma_+(-\boldsymbol{\Omega},\mathbf{r}) = \int d\Omega_L \eta(-\boldsymbol{\Omega}_L,\mathbf{r})|\boldsymbol{\Omega}\cdot\boldsymbol{\Omega}_L|_+ \tag{29}$$

**$\varphi_L$- symmetric leaf area densities**: Let us now consider the case that the leaf area densities depend only on the *inclination* $\mu_L$ of leaves (hence are $\varphi_L$-symmetric), that is:

$$\eta(\boldsymbol{\Omega}_L,\mathbf{r}) = \tfrac{1}{2\pi}\tilde{\eta}(\mu_L,\mathbf{r}) = \tfrac{1}{2\pi}\eta(\mathbf{r})\tilde{\lambda}(\mu_L,\mathbf{r}) \tag{30}$$

(see (11) and (12)). We then get from (28), using (4) and (14):

$$\Delta\Gamma(\boldsymbol{\Omega},\mathbf{r}) = \Delta\Gamma(\mu,\mathbf{r}) = \tfrac{1}{2\pi}\int_{-1}^{1} d\mu_L \tilde{\eta}(\mu_L,\mathbf{r})\int_{-\pi}^{\pi} d\varphi_L\, \boldsymbol{\Omega}\cdot\boldsymbol{\Omega}_L = \eta(\mathbf{r})\mu\langle\mu_L\rangle_{\mathbf{r}}$$

$$\Gamma(\boldsymbol{\Omega},\mathbf{r}) = \Gamma(\mu,\mathbf{r}) = \tfrac{1}{2\pi}\int_{-1}^{1} d\mu_L \tilde{\eta}(\mu_L,\mathbf{r})\int_{-\pi}^{\pi} d\varphi_L\, |\boldsymbol{\Omega}\cdot\boldsymbol{\Omega}_L| = \eta(\mathbf{r})\langle g(\mu,\mu_L)\rangle_{\mathbf{r}} \tag{31}$$

where $\langle\mu_L\rangle_{\mathbf{r}}$ is the mean leaf inclination at $\mathbf{r}$, and the function $g(\mu,\mu')$ was defined in (6). Note, as is intuitively obvious, that $\Gamma(\pm\mathbf{u}_z,\mathbf{r}) = \eta(\mathbf{r})\langle\hat{\mu}_L\rangle$ since $g(\pm 1,\mu_L) = \hat{\mu}_L$. The coefficients (31) have simple analytic expressions for the following distributions of leaf inclinations:

*Horizontal*: $\tilde{\lambda}(\mu_L,\mathbf{r}) = \delta(\mu_L - 1)$

$$\Gamma_\pm(\boldsymbol{\Omega},\mathbf{r}) = \eta(\mathbf{r})\hat{\mu}\,\Theta(\pm\mu), \qquad \Gamma(\boldsymbol{\Omega},\mathbf{r}) = \eta(\mathbf{r})\hat{\mu} \tag{32}$$

*Erect*: $\tilde{\lambda}(\mu_L,\mathbf{r}) = \delta(\mu_L)$

$$\Gamma_\pm(\boldsymbol{\Omega},\mathbf{r}) = \tfrac{1}{\pi}\eta(\mathbf{r})\sin\theta, \qquad \Gamma(\boldsymbol{\Omega},\mathbf{r}) = \tfrac{2}{\pi}\eta(\mathbf{r})\sin\theta \tag{33}$$

*Isotropic*: $\tilde{\lambda}(\mu_L,\mathbf{r}) = \tfrac{1}{2}$

$$\Gamma_\pm(\boldsymbol{\Omega},\mathbf{r}) = \tfrac{1}{4}\eta(\mathbf{r}), \qquad \Gamma(\boldsymbol{\Omega},\mathbf{r}) = \tfrac{1}{2}\eta(\mathbf{r}) \tag{34}$$

*Semi-isotropic*: $\tilde{\lambda}(\mu_L,\mathbf{r}) = \Theta(\mu_L)$

$$\Gamma(\boldsymbol{\Omega},\mathbf{r}) = \tfrac{1}{2}\eta(\mathbf{r}), \qquad \Delta\Gamma(\boldsymbol{\Omega},\mathbf{r}) = \tfrac{1}{2}\eta(\mathbf{r})\mu \tag{35}$$

The first two cases are obvious, since $\boldsymbol{\Omega}\cdot\boldsymbol{\Omega}_L = \mu$ for horizontal leaves, and $\boldsymbol{\Omega}\cdot\boldsymbol{\Omega}_L = \sin\theta \geq 0$ for erect leaves. The isotropic case follows from (4). In the semi-isotropic case, one gets $\Delta\Gamma$ from (31), and $\Gamma$ from (29) using $\Theta(\mu_L) + \Theta(-\mu_L) = 1$ and (4). In the erect and isotropic cases, $\Gamma_+ = \Gamma_- = \tfrac{1}{2}\Gamma$ since these leaf orientation distributions destroy the distinction between leaf tops and leaf bottoms (the mean leaf inclination is zero).

## 5 Horizontal homogeneity

We will now assume that all quantities are independent of the horizontal coordinates $x, y$. Also, we let the canopy extend vertically from the ground at $z = z_g$ to its top at $z_0$. Thus, the leaf area densities $\eta(\Omega_L, \mathbf{r}) = \eta(\Omega_L, z)$ vary vertically only, and vanish outside $(z_0, z_g)$.

*Leaf area index* (LAI): With the total leaf area density $\eta(\mathbf{r}) = \eta(z)$ varying vertically only, it is useful to define the *leaf area index* (LAI) $\mathcal{L}(z_1, z_2)$ of a horizontal layer $(z_1, z_2)$ as:

$$\mathcal{L}(z_1, z_2) \equiv \left| \int_{z_1}^{z_2} dz\, \eta(z) \right| = \text{total leaf area between } z_1 \text{ and } z_2 \text{ per unit horizontal area.} \quad (36)$$

*Vertical attenuation*: The probability $\xi_\Omega(z_1, z_2)$ for a photon $\Omega$ to pass straight through a horizontal layer $(z_1, z_2)$ is given by (26)-(27) with $d\ell = dz/\mu$:

$$\xi_\Omega(z_1, z_2) = e^{-|\Lambda_\Omega(z_1, z_2)|}, \qquad \Lambda_\Omega(z_1, z_2) \equiv \int_{z_1}^{z_2} dz\, \Gamma(\Omega, z)/\mu \quad (37)$$

Note that if all the leaves are *horizontal*, then the fraction of photons $\Omega$ intercepted by a horizontal layer $dz$, namely $\Gamma(\Omega, z) dz/\hat{\mu} = \eta(z) dz$ by (32), is independent of $\hat{\mu}$ (the distance travelled is $d\ell = dz/\hat{\mu}$, but the 'view factor' $|\Omega \cdot \Omega_L| = \hat{\mu}$); then, $\hat{\Lambda}_\Omega(z_1, z_2) = \mathcal{L}(z_1, z_2)$ is just the LAI, and $\xi_\Omega(z_1, z_2) = e^{-\mathcal{L}(z_1, z_2)}$ is independent of $\Omega$. But in general, $\xi_\Omega(z_1, z_2)$ depends strongly on the inclination $\hat{\mu}$ of light rays. For example, $\hat{\Lambda}_\Omega(z_1, z_2) = \tfrac{1}{2}\mathcal{L}(z_1, z_2)/\hat{\mu}$ in the semi-isotropic case (35). Thus, the LAI alone is not in general a reliable indicator of optical thickness.

*Down and up fluxes*: We will write

$$I(\Omega, z) = \begin{cases} D(\Omega, z) & \text{if } \Omega \in D \quad (\textit{i.e.,} \text{ if } \mu > 0) \\ U(\Omega, z) & \text{if } \Omega \in U \quad (\textit{i.e.,} \text{ if } \mu < 0) \end{cases} \quad (38)$$

where $\Omega \in D$ or $\Omega \in U$ are shorthands for '$\Omega$ points downwards', or 'upwards'. Also, we denote by $\mathcal{D}(z)$ and $\mathcal{U}(z)$ the *total* 'down' and 'up' vertical photon fluxes, i.e., the numbers of photons crossing a *horizontal* unit area at height $z$, per second:

$$\mathcal{D}(z) \equiv \int_{\mu>0} d\Omega\, \hat{\mu}\, D(\Omega, z), \qquad \mathcal{U}(z) \equiv \int_{\mu<0} d\Omega\, \hat{\mu}\, U(\Omega, z) \quad (39)$$

We will sometimes use the notations $I_0(\Omega) \equiv I(\Omega, z_0)$, $\mathcal{D}_g \equiv \mathcal{D}(z_g)$, etc.

*The LTE*: With all quantities dependent on the vertical coordinate $z$ only, one can replace everywhere $\mathbf{r}$ by $z$. Also, $\Omega \cdot \nabla = \mu\, \partial/\partial z$, so that the LTE (22) becomes

$$\mu_f \frac{d}{dz} I(\Omega_f, z) = -\Gamma(\Omega_f, z) I(\Omega_f, z) + \int d\Omega_i \, I(\Omega_i, z) S(\Omega_i \to \Omega_f, z) + \mathcal{E}(\Omega_f, z) \quad (40)$$

The continuity equation (23) becomes, since $\nabla \cdot \mathbf{i} = \partial_x i_x + \partial_y i_y + \partial_z i_z$, where $\partial_x \equiv \partial/\partial x$, etc.:

$$\frac{d}{dz} i(z) = \mathcal{E}(z) - \mathcal{A}(z), \qquad i(z) \equiv i_z(z) = \mathcal{D}(z) - \mathcal{U}(z) \quad (41)$$

***Black leaves***: Leaves which do not scatter any light (absorbing it all), and moreover emit no light, will be called 'black' leaves. If all the leaves are black, then $S(\Omega_i \to \Omega_f, z) = 0$ and $\mathcal{E}(\Omega_f, z) = 0$, in which case (40) has the solution

$$I(\Omega, z) = I(\Omega, z_0) e^{-\Lambda_\Omega(z_0, z)}, \qquad \Lambda_\Omega(z_0, z) = \int_{z_0}^{z} dz' \, \Gamma(\Omega, z')/\mu \quad (42)$$

Here, 'down' radiances ($\mu > 0$) decrease exponentially, while 'up' radiances ($\mu < 0$) increase exponentially, as we go from the top $z_0$ to the bottom $z_g$ of the canopy.

## 6  Ground boundary conditions, energy balance

We now specify the boundary conditions at the ground surface $z = z_g$, and then state the global energy balance condition.

***The ground boundary***: Horizontally homogeneous ground optical coefficients $\sigma_g(\Omega_i \to \Omega_f)$, $\alpha_g(\Omega_i)$ and $\varepsilon_g(\Omega_f)$, are defined as for a horizontal leaf, but only for $\Omega_i \in D$ and $\Omega_f \in U$, since the ground receives only 'down' light, and 'emits' only 'up' light. Similarly to (16),

$$\alpha_g(\Omega_i) = 1 - \int_{\mu_f < 0} d\Omega_f \, \sigma_g(\Omega_i \to \Omega_f) \qquad (\Omega_i \in D) \quad (43)$$

Since a unit area of ground intercepts from direction $\Omega_i$, and emits in direction $\Omega_f$, beams of cross-sections $\hat{\mu}_i$ and $\hat{\mu}_f$, the 'up' radiance just above the ground is

$$U(\Omega_f, z_g) = \int_{\mu_i > 0} d\Omega_i \, D(\Omega_i, z_g) \left( \hat{\mu}_i / \hat{\mu}_f \right) \sigma_g(\Omega_i \to \Omega_f) + \hat{\mu}_f^{-1} \varepsilon_g(\Omega_f) \qquad (\mu_f < 0) \quad (44)$$

The scattering term may be viewed as due to an 'emission' $\varepsilon_g^{scatt}(\Omega_f)$, similarly to (21). Multiplying (44) by $\hat{\mu}_f$, and integrating over $\Omega_f \in U$, yields the 'up' current $\mathcal{U}(z_g)$. Substracting $\mathcal{U}(z_g)$ from the 'down' current $\mathcal{D}(z_g) = \int_{\mu_i > 0} d\Omega_i \, D(\Omega_i, z_g) \hat{\mu}_i$, and using (43), yields the net vertical current just above the ground:

$$i(z_g) = \mathcal{D}(z_g) - \mathcal{U}(z_g) = \mathcal{A}_g - \mathcal{E}_g$$

$$\mathcal{E}_g \equiv \int_{\mu_f < 0} d\Omega_f \, \varepsilon_g(\Omega_f), \qquad \mathcal{A}_g \equiv \int_{\mu_i > 0} d\Omega_i \, D(\Omega_i, z_g) \hat{\mu}_i \alpha_g(\Omega_i) \quad (45)$$

where $\mathcal{E}_g$ and $\mathcal{A}_g$ are the total photon emission and absorption rates per unit area of ground.

***Energy balance***: Energy conservation requires that, at the top $z = z_0$ of the canopy:

$$\mathcal{D}(z_0) - \mathcal{U}(z_0) = \hat{\mathcal{A}} - \hat{\mathcal{E}}$$

$$\hat{\mathcal{E}} \equiv \int_{z_0}^{z_g} dz \, \mathcal{E}(z) + \mathcal{E}_g, \qquad \hat{\mathcal{A}} \equiv \int_{z_0}^{z_g} dz \, \mathcal{A}(z) + \mathcal{A}_g \quad (46)$$

where $\hat{\mathcal{E}}$ and $\hat{\mathcal{A}}$ are the total emission and absorption rates of the whole canopy plus ground, per unit horizontal area. If $\hat{\mathcal{E}} = \hat{\mathcal{A}}$, then $\mathcal{U}_0 = \mathcal{D}_0$ (what goes in must come out). Note that if $\mathcal{E}(z) = \mathcal{A}(z)$ at all heights $z$, then $\mathcal{D}(z) - \mathcal{U}(z) = \mathcal{A}_g - \mathcal{E}_g$ throughout the canopy, due the continuity equation (41). If $\mathcal{E}(z) = \mathcal{A}(z)$ at all $z$, and moreover $\mathcal{E}_g = \mathcal{A}_g$, then $\mathcal{U}(z) = \mathcal{D}(z)$ at all $z$. This is the case, e.g., with non-emitting non-absorbing 'white' leaves and ground. All these relations provide very useful tests.

## 7  First-scattered sunlight treated as an 'emission'

To do numerical computations, one must discretize photon directions. Because sharply directional sunlight is very intense, putting its direction equal to one of the discrete directions may entail sizable errors. So it is customary [1-3] to compute accurately the sunlight which has been scattered *once*, and treat that as an 'emission'.

***Incident sunlight and skylight***: Sunlight incident on the top $z_0$ of the canopy will be modeled as a sharply directional radiance $D_0^h(\Omega) = h_0 \delta(\Omega - \Omega_h)$, where $\Omega_h$ is the (downwards) direction of sun rays ($h$ stands for 'helios'). The incident diffuse skylight will be assumed 'down' isotropic, i.e., of the same radiance in all 'down' directions, $D_0^d(\Omega) = D_0^d$ ($d$ stands for *diffuse*). Thus, the total radiance $D_0(\Omega) = D(\Omega, z_0)$ incident on the top $z_0$ of the canopy is

$$D(\Omega, z_0) = D_0^h(\Omega) + D_0^d(\Omega)$$

$$D_0^h(\Omega) = h_0 \delta(\Omega - \Omega_h), \qquad D_0^d(\Omega) = D_0^d \quad (47)$$

The incident 'down' *vertical* fluxes will be denoted $\mathcal{D}_0^h$ and $\mathcal{D}_0^d$, that is, by (4) and (5):

$$\mathcal{D}_0^h = \int_{\mu > 0} d\Omega \, \hat{\mu} \, D_0^h(\Omega) = h_0 \hat{\mu}_h, \qquad \mathcal{D}_0^d = \pi D_0^d \quad (48)$$

These fluxes are assumed known (i.e., measured experimentally).

**Sunlight inside the canopy**: Inside the canopy, the sunlight radiance $D^h(\Omega,z)$, and vertical flux $\mathcal{D}^h(z)$, are attenuated due to interception by leaves. Writing $\mathcal{D}^h(z) = h(z)\hat{\mu}_h$, so that $h(z_0) = h_0$ in view of (48), we have:

$$D^h(\Omega,z) = h(z)\,\delta(\Omega-\Omega_h), \qquad \mathcal{D}^h(z) = h(z)\hat{\mu}_h$$
$$h(z) = h_0\,\xi_h(z), \qquad \xi_h(z) \equiv \xi_{\Omega_h}(z_0,z) = e^{-\int_{z_0}^{z}\Gamma(\Omega_h,z')dz'/\hat{\mu}_h} \tag{49}$$

Since the attenuation $\xi_h(z)$ depends sensitively on $\hat{\mu}_h$, in general, sizable errors may ensue from putting $\Omega_h$ equal to one of the discretized photon directions, as said earlier.

**LTE for the diffuse radiance**: Separating the total radiance $I(\Omega,z)$ into 'sun' and 'diffuse' components, $I = I^d + D^h$, with $D^h(\Omega,z)$ given by (49), and substituting into (40) and (44), we get the LTE for the diffuse radiance:

$$\mu_f \frac{d}{dz}I^d(\Omega_f,z) = -\Gamma(\Omega_f,z)I^d(\Omega_f,z) + \int d\Omega_i\, I^d(\Omega_i,z)S(\Omega_i\to\Omega_f,z) + \mathcal{E}^{tot}(\Omega_f,z)$$
$$U^d(\Omega_f,z_g) = \int_{\mu_i>0} d\Omega_i\, D^d\left(\hat{\mu}_i/\hat{\mu}_f\right)\sigma_g(\Omega_i\to\Omega_f) + \hat{\mu}_f^{-1}\varepsilon_g^{tot}(\Omega_f) \tag{50}$$

where $\mathcal{E}^{tot} = \mathcal{E} + \mathcal{E}^h$ and $\varepsilon_g^{tot} = \varepsilon_g + \varepsilon_g^h$ are sums of 'true' emissions, $\mathcal{E}$ and $\varepsilon_g$, and of first-scattered-sunlight 'emissions'

$$\mathcal{E}^h(\Omega_f,z) = h(z)S(\Omega_h\to\Omega_f,z), \qquad \varepsilon_g^h(\Omega_f) = \mathcal{D}^h(z_g)\sigma_g(\Omega_h\to\Omega_f) \tag{51}$$

Note, refering to (20), that one may define first-scattered sunlight leaf 'emission' coefficients

$$\varepsilon_{\Omega_L}^h(\Omega_f,z) = h(z)|\Omega_h\cdot\Omega_L|\sigma_{\Omega_L}(\Omega_h\to\Omega_f) \tag{52}$$

The vertical fluxes are $\mathcal{D}(z) = \mathcal{D}^d(z) + \mathcal{D}^h(z)$ and $\mathcal{U}(z) = \mathcal{U}^d(z)$, where $\mathcal{D}^d$ and $\mathcal{U}^d$ are given by (39) with $I^d$ instead of $I$. The canopy absorption rate $\mathcal{A}(z)$ may be written, on substituting $I = I^d + D^h$ into (24):

$$\mathcal{A} = \mathcal{A}^d + \mathcal{A}^h$$
$$\mathcal{A}^d(z) = \int d\Omega\, I^d(\Omega,z)A(\Omega,z), \qquad \mathcal{A}^h(z) = h(z)A(\Omega_h,z) \tag{53}$$

## Part II  Lambertian scattering and emission

The above completes the general aspects of the theory. We now specialize to Lambertian scattering and emission. Section 8 first defines Lambertian leaf and ground optical coefficients. Section 9 then computes the canopy local optical coefficients for various azymuthally symmetric (i.e., $\varphi_L$-symmetric) distributions of leaf orientations.

## 8  Lambertian leaves and ground

A rugged horizontal surface tends to emit equal radiances in all 'up' directions. Such 'up' isotropic emission is called Lambertian. Since a unit horizontal area emits a beam of cross-section $\hat{\mu}$ in direction $\Omega$, the *number* of photons $\Omega$ it emits is proportional to $\hat{\mu}$.

*Lambertian leaf optical coefficients*: Thus, for a *horizontal* Lambertian leaf, we set

$$\sigma(\Omega_i \to \Omega_f) = \tfrac{1}{\pi}\hat{\mu}_f\,\hat{\sigma}_{\operatorname{sgn}\mu_i,\operatorname{sgn}\mu_f}, \qquad \varepsilon(\Omega_f) = \tfrac{1}{\pi}\hat{\mu}_f\,\hat{\varepsilon}_{\operatorname{sgn}\mu_f} \tag{54}$$

($\operatorname{sgn}\mu = +$ or $-$ is the sign of $\mu$) where we allow the top and bottom sides of the leaf to have different optical properties. We shall also use the notations $\hat{\sigma}_{dd} \equiv \hat{\sigma}_{++}$, $\hat{\sigma}_{du} \equiv \hat{\sigma}_{+-}$, etc., where $d$ stands for 'down', and $u$ for 'up'. Since $\tfrac{1}{\pi}\int_{\pm\mu>0}\hat{\mu}\,d\Omega = 1$, we see that $\hat{\varepsilon}_+$ ($\hat{\varepsilon}_-$) is the total numbers of photons emitted, per second per unit area, by the top (bottom) of the leaf. Similarly, $\hat{\sigma}_{dd} \equiv \hat{\sigma}_{++}$ ($\hat{\sigma}_{uu} \equiv \hat{\sigma}_{--}$) is the fraction of 'down' ('up') light transmitted through the leaf, and $\hat{\sigma}_{du} \equiv \hat{\sigma}_{+-}$ ($\hat{\sigma}_{ud} \equiv \hat{\sigma}_{-+}$) the fraction reflected off the top (bottom) of the leaf. The fractions absorbed are, by (16):

$$\alpha(\Omega_i) = \begin{cases} \hat{\alpha}_+ \equiv 1 - \hat{\sigma}_{++} - \hat{\sigma}_{+-} & \text{if } \mu_i > 0 \\ \hat{\alpha}_- \equiv 1 - \hat{\sigma}_{-+} - \hat{\sigma}_{--} & \text{if } \mu_i < 0 \end{cases} \qquad (0 \le \hat{\alpha}_\pm \le 1) \tag{55}$$

Note the constraints $0 \le \hat{\sigma}_{\operatorname{sgn}\mu_i,\operatorname{sgn}\mu_f} \le 1$, and $\hat{\sigma}_{\pm+} + \hat{\sigma}_{\pm-} \le 1$. It is convenient to also denote

$$\hat{\alpha} \equiv \hat{\alpha}_+ + \hat{\alpha}_-, \quad \Delta\hat{\alpha} \equiv \hat{\alpha}_+ - \hat{\alpha}_-, \qquad \hat{\varepsilon} \equiv \hat{\varepsilon}_+ + \hat{\varepsilon}_-, \quad \Delta\hat{\varepsilon} \equiv \hat{\varepsilon}_+ - \hat{\varepsilon}_- \tag{56}$$

*Lambertian ground optical coefficients*: For the ground coefficients, we set

$$\sigma_g(\Omega_i \to \Omega_f) = \tfrac{1}{\pi}\hat{\mu}_f\,\hat{\sigma}_g, \qquad \varepsilon_g(\Omega_f) = \tfrac{1}{\pi}\hat{\mu}_f\,\hat{\varepsilon}_g, \qquad \hat{\alpha}_g = 1 - \hat{\sigma}_g \tag{57}$$

(for $\mu_i > 0$ and $\mu_f < 0$), where $\hat{\varepsilon}_g$ is the total emission rate, and $\hat{\sigma}_g$ the fraction of light reflected. Equation (45) then becomes

$$\mathcal{D}_g - \mathcal{U}_g = \widehat{\alpha}_g \mathcal{D}_g - \widehat{\varepsilon}_g \quad \text{or} \quad \mathcal{U}_g = \widehat{\sigma}_g \mathcal{D}_g + \widehat{\varepsilon}_g \tag{58}$$

In view of (51), the first-scattered-sunlight 'emission' by the ground is $\widehat{\varepsilon}_g^h = \widehat{\sigma}_g \mathcal{D}^h(z_g)$.

***Inclined leaves***: For inclined leaves with normals $\Omega_L$, terms $\mu = \Omega \cdot \mathbf{u}_z$ in (54) become $\Omega \cdot \Omega_L$:

$$\sigma_{\Omega_L}(\Omega_i \to \Omega_f) = \tfrac{1}{\pi}|\Omega_i \cdot \Omega_L|\widehat{\sigma}_{\mathrm{sgn}\,\Omega_i \cdot \Omega_L,\,\mathrm{sgn}\,\Omega_f \cdot \Omega_L}$$
$$\varepsilon_{\Omega_L}(\Omega_f,\mathbf{r}) = \tfrac{1}{\pi}|\Omega_f \cdot \Omega_L|\widehat{\varepsilon}_{\Omega_L,\,\mathrm{sgn}\,\Omega_f \cdot \Omega_L}(\mathbf{r}), \qquad \widehat{\varepsilon}_{\Omega_L,\pm}^h(\mathbf{r}) = h(\mathbf{r})|\Omega_h \cdot \Omega_L|\widehat{\sigma}_{\mathrm{sgn}\,\Omega_h \cdot \Omega_L,\pm} \tag{59}$$

where we let $\widehat{\varepsilon}_{\Omega_L,\pm}(\mathbf{r})$ depend on $\Omega_L$, as well as on $\mathbf{r}$, since the temperature of a leaf depends in general on its orientation (especially relative to sun rays), and so does the first-scattered-sunlight 'emission' coefficient $\widehat{\varepsilon}_{\Omega_L,\pm}^h(\mathbf{r})$, following from (52).

***Canopy optical coefficients***: Using (1) and (18), we obtain for the canopy scattering coefficient:

$$S(\Omega_i \to \Omega_f, \mathbf{r}) = \tfrac{1}{\pi}\int d\Omega_L\, \eta(\Omega_L,\mathbf{r}) \sum_{u_i,u_f=\pm} \widehat{\sigma}_{u_i,u_f} |\Omega_i \cdot \Omega_L|_{u_i} |\Omega_f \cdot \Omega_L|_{u_f} \tag{60}$$

The integral of (60) over $\Omega_f$ yields $\sum_{u_i,u_f} \widehat{\sigma}_{u_i,u_f} \Gamma_{u_i}(\Omega_i,\mathbf{r})$, by (28), or using (55)-(56):

$$\int d\Omega_f\, S(\Omega_i \to \Omega_f, \mathbf{r}) = (\widehat{\sigma}_{++} + \widehat{\sigma}_{+-})\Gamma_+(\Omega_i,\mathbf{r}) + (\widehat{\sigma}_{--} + \widehat{\sigma}_{-+})\Gamma_-(\Omega_i,\mathbf{r})$$
$$= \Gamma(\Omega_i,\mathbf{r}) - A(\Omega_i,\mathbf{r}) \tag{61}$$

$$A(\Omega,\mathbf{r}) = \widehat{\alpha}_+ \Gamma_+(\Omega,\mathbf{r}) + \widehat{\alpha}_- \Gamma_-(\Omega,\mathbf{r}) = \tfrac{1}{2}\widehat{\alpha}\,\Gamma(\Omega,\mathbf{r}) + \tfrac{1}{2}\Delta\widehat{\alpha}\,\Delta\Gamma(\Omega,\mathbf{r}) \tag{62}$$

in accordance with (19). All these results are quite intuitive. The local canopy emission $\mathcal{E}(\Omega,\mathbf{r})$ in direction $\Omega$, and the total local emission $\mathcal{E}(\mathbf{r})$, per unit volume at $\mathbf{r}$, are

$$\mathcal{E}(\Omega,\mathbf{r}) = \tfrac{1}{\pi}\int d\Omega_L\, \eta(\Omega_L,\mathbf{r})\left[\widehat{\varepsilon}_{\Omega_L,+}(\mathbf{r})|\Omega \cdot \Omega_L|_+ + \widehat{\varepsilon}_{\Omega_L,-}(\mathbf{r})|\Omega \cdot \Omega_L|_-\right] \tag{63}$$

$$\mathcal{E}(\mathbf{r}) = \int d\Omega\, \mathcal{E}(\Omega,\mathbf{r}) = \int d\Omega_L\, \eta(\Omega_L,\mathbf{r})\widehat{\varepsilon}_{\Omega_L}(\mathbf{r}), \qquad \widehat{\varepsilon}_{\Omega_L}(\mathbf{r}) \equiv \widehat{\varepsilon}_{\Omega_L,+}(\mathbf{r}) + \widehat{\varepsilon}_{\Omega_L,-}(\mathbf{r}) \tag{64}$$

If $\widehat{\varepsilon}_{\Omega_L,\pm}(\mathbf{r}) = \widehat{\varepsilon}_\pm(\mathbf{r})$ is independent of $\Omega_L$, then $\mathcal{E}(\mathbf{r}) = \eta(\mathbf{r})\widehat{\varepsilon}(\mathbf{r})$, and, on using (28):

$$\mathcal{E}(\Omega,\mathbf{r}) = \tfrac{1}{\pi}\widehat{\varepsilon}_+(\mathbf{r})\Gamma_+(\Omega,\mathbf{r}) + \tfrac{1}{\pi}\widehat{\varepsilon}_-(\mathbf{r})\Gamma_-(\Omega,\mathbf{r})$$
$$= \tfrac{1}{2\pi}\widehat{\varepsilon}(\mathbf{r})\Gamma(\Omega,\mathbf{r}) + \tfrac{1}{2\pi}\Delta\widehat{\varepsilon}(\mathbf{r})\Delta\Gamma(\Omega,\mathbf{r}) \qquad (\text{if } \widehat{\varepsilon}_{\Omega_L,\pm}(\mathbf{r}) = \widehat{\varepsilon}_\pm(\mathbf{r})) \tag{65}$$

## 9  Canopy optical coefficients for $\varphi_L$- symmetric Lambertian leaves

Let us now calculate the Lambertian canopy optical coefficients in the case that the leaf area density $\eta(\Omega_L,\mathbf{r}) = \frac{1}{2\pi}\tilde{\eta}(\mu_L,\mathbf{r})$ is $\varphi_L$-symmetric. Using (8) and (14), we get from (60):

$$S(\Omega_i \to \Omega_f,\mathbf{r}) = \eta(\mathbf{r})\sum_{u_i,u_f=\pm}\hat{\sigma}_{u_i,u_f}\left\langle g_{u_i,u_f}(\mu_i,\mu_f,\mu_L,\varphi_{if})\right\rangle_{\mathbf{r}} \quad (\varphi_{if} \equiv \varphi_i - \varphi_f) \tag{66}$$

where $g_{u_i,u_f}(\mu_i,\mu_f,\mu_L,\varphi_{if})$ is known analytically, and the average is over $\mu_L$.

In the case of horizontal, isotropic, and erect (Lambertian) leaves, $S(\Omega_i \to \Omega_f,\mathbf{r})$ has relatively simple analytic expressions. These are useful to indicate the allure of these coefficients whenever a real canopy approaches one of these cases. If moreover

$$\hat{\varepsilon}_{\Omega_L\pm}(\mathbf{r}) = \hat{\varepsilon}_\pm(\mathbf{r}) \quad \text{is independent of } \Omega_L \tag{67}$$

as will be assumed in the rest of this section, then $\mathcal{E}(\Omega,\mathbf{r})$ is immediate from (65) and the analytic expressions (32)-(35) for $\Gamma_\pm$. For instance, in the semi-isotropic case (35) (a case for which $S(\Omega_i \to \Omega_f,\mathbf{r})$ is *not* known analytically), one gets, assuming (67):

$$\mathcal{E}(\Omega,\mathbf{r}) = \tfrac{1}{4\pi}\eta(\mathbf{r})[\hat{\varepsilon}(\mathbf{r}) + \mu\Delta\hat{\varepsilon}_+(\mathbf{r})] \tag{68}$$

***Horizontal leaves***: Putting $\eta(\Omega_L,\mathbf{r}) = \tfrac{1}{2\pi}\eta(\mathbf{r})\delta(\mu_L-1)$ in (18) and (20), yields:

$$\begin{aligned}S(\Omega_i \to \Omega_f,\mathbf{r}) &= \eta(\mathbf{r})\hat{\mu}_i\,\sigma(\Omega_i \to \Omega_f) = \tfrac{1}{\pi}\eta(\mathbf{r})\hat{\mu}_i\hat{\mu}_f\hat{\sigma}_{\mathrm{sgn}\,\mu_i,\mathrm{sgn}\,\mu_f}\\ \mathcal{E}(\Omega,\mathbf{r}) &= \eta(\mathbf{r})\varepsilon(\Omega,\mathbf{r}) = \tfrac{1}{\pi}\eta(\mathbf{r})\hat{\mu}\,\hat{\varepsilon}_{\mathrm{sgn}\,\mu}(\mathbf{r})\end{aligned} \tag{69}$$

if (67) holds. Thus, a horizontal layer $dz$ of canopy acts as a single large horizontal leaf.

***Erect leaves***: With $\eta(\Omega_L,\mathbf{r}) = \tfrac{1}{2\pi}\eta(\mathbf{r})\delta(\mu_L)$, we get from (60),(65),(67) and (33), using (B.14) and $\zeta(\mu)$ defined in (7):

$$\begin{aligned}S(\Omega_i \to \Omega_f,\mathbf{r}) &= \tfrac{1}{4\pi^2}\eta(\mathbf{r})\sin\theta_i\sin\theta_f\left[(\hat{\sigma}_{++}+\hat{\sigma}_{--})\zeta(-\cos\varphi_{if}) + (\hat{\sigma}_{+-}+\hat{\sigma}_{-+})\zeta(\cos\varphi_{if})\right]\\ \mathcal{E}(\Omega,\mathbf{r}) &= \pi^{-2}\eta(\mathbf{r})\hat{\varepsilon}(\mathbf{r})\sin\theta\end{aligned} \tag{70}$$

where $\varphi_{if} \equiv \varphi_i - \varphi_f$. These, like $\Gamma_\pm(\Omega,\mathbf{r})$ in (33), vanish in vertical directions (as expected since $\sin\theta$ plays, for vertical leaves, the role of $\mu = \cos\theta$ for horizontal leaves).

***Isotropic leaves***: With $\eta(\Omega_L,\mathbf{r}) = \tfrac{1}{4\pi}\eta(\mathbf{r})$, there is no prefered direction, so that $S(\Omega_i \to \Omega_f,\mathbf{r})$ is a function only of the angle between $\Omega_i$ and $\Omega_f$, or equivalently of its cosine $\Omega_i\cdot\Omega_f$. Likewise, $\Gamma(\Omega,\mathbf{r})$ is isotropic, see (34), and so is the emission $\mathcal{E}(\Omega,\mathbf{r})$ if (67) holds. We indeed get from (60),(65),(67) and (34), using (B.13):

$$S(\Omega_i \to \Omega_f, \mathbf{r}) = \tfrac{1}{3\pi}\eta(\mathbf{r})\left[(\hat{\sigma}_{++} + \hat{\sigma}_{--})\zeta(-\Omega_i \cdot \Omega_f) + (\hat{\sigma}_{+-} + \hat{\sigma}_{-+})\zeta(\Omega_i \cdot \Omega_f)\right]$$
$$\mathcal{E}(\Omega, \mathbf{r}) = \tfrac{1}{4\pi}\eta(\mathbf{r})\hat{\varepsilon}(\mathbf{r})$$
(71)

Since $\zeta(\mu)$ decreases smoothly from $\zeta(-1) = \pi$ to $\zeta(0) = 1$ to $\zeta(1) = 0$, the function $S(\Omega_i \to \Omega_f, \mathbf{r})$ in (71) describes scattering that is predominantly backwards if the transmissions $\hat{\sigma}_{++} = \hat{\sigma}_{--} = 0$, and forwards if the reflections $\hat{\sigma}_{+-} = \hat{\sigma}_{-+} = 0$. Note that scattering is anisotropic even if $\hat{\sigma}_{++} = \hat{\sigma}_{+-} = \hat{\sigma}_{-+} = \hat{\sigma}_{--}$, being minimum in the sideways directions, because it is the leaves nearly perpendicular to light rays which capture the most light (due $\Omega_i \cdot \Omega_L$), and these leaves in turn 're-emit' preferentially along their normals (due $\Omega_f \cdot \Omega_L$).

**Part III  Analytically solvable models**

We will now seek situations permitting the light climate $I(\Omega, \mathbf{r})$ to be calculated analytically. In section 10 we find conditions leading to an isotropic light climate $I(\Omega, \mathbf{r}) = I(\mathbf{r})$ (equal radiances in all directions), in which case the LTE simplifies and can be solved analytically. Section 11 deals with horizontal leaves, for which the LTE can again be solved analytically. Finally, section 12 studies a case of extreme light trapping by horizontal leaves.

***Black leaves***: We first note a simple solvable case: With totally absorbing, non-emitting 'black' leaves, and a distribution of leaf orientations $\lambda(\Omega_L, z) = \tfrac{1}{2\pi}\tilde{\lambda}(\mu_L)$ uniform in $\varphi_L$ and in $z$, the radiances inside the canopy are, by (42) and (31):

$$I(\Omega, z) = I(\Omega, z_0)e^{-\mathcal{L}(z)\langle g(\mu, \mu_L)\rangle}, \qquad \mathcal{L}(z) \equiv \mathcal{L}(z_0, z) \tag{72}$$

where $\mathcal{L}(z)$ is the total LAI above height $z$, see (36).

**10  Cases for which the radiance is isotropic**

In this section, we find conditions producing isotropic radiances. This is of interest because putting $I(\Omega, \mathbf{r}) = I(\mathbf{r})$ into the LTE (22),(44) yields the far simpler LTE

$$\Omega_f \cdot \nabla I(\mathbf{r}) = -\Gamma(\Omega_f, \mathbf{r})I(\mathbf{r}) + I(\mathbf{r})\int d\Omega_i\, S(\Omega_i \to \Omega_f, \mathbf{r}) + \mathcal{E}(\Omega_f, \mathbf{r})$$
$$= \int d\Omega_L\, \eta(\Omega_L, \mathbf{r})\left\{|\Omega_f \cdot \Omega_L|\left(b_{\Omega_L}(\Omega_f) - 1\right)I(\mathbf{r}) + \varepsilon_{\Omega_L}(\Omega_f, \mathbf{r})\right\}$$
(73)

$$I(z_g) = b_g(\Omega_f)I(z_g) + \hat{\mu}_f^{-1}\varepsilon_g(\Omega_f) \tag{74}$$

where we used (18) and (20), and defined

$$b_{\Omega_L}(\Omega_f) \equiv |\Omega_f \cdot \Omega_L|^{-1} \int d\Omega_i |\Omega_i \cdot \Omega_L| \sigma_{\Omega_L}(\Omega_i \to \Omega_f)$$
$$b_g(\Omega_f) \equiv \hat{\mu}_f^{-1} \int_{\Omega_i \in D} d\Omega_i \, \hat{\mu}_i \, \sigma_g(\Omega_i \to \Omega_f) \tag{75}$$

Imagine now that an infinitesimal leaf element is inserted at $\mathbf{r}$ amid an isotropic light climate $I(\Omega,\mathbf{r}) = I(\mathbf{r})$. The latter will not be disturbed if each side of the leaf 'emits' a radiance identical to that intercepted. Similarly for the ground surface. So we must have, for any direction $\Omega_f$ (in the following, the 'ground' equations apply only to the horizontally homogeneous case):

$$b(\Omega_f)I(\mathbf{r}) + \hat{\mu}_f^{-1}\varepsilon(\Omega_f,\mathbf{r}) = I(\mathbf{r}), \qquad b_g(\Omega_f)I(z_g) + \hat{\mu}_f^{-1}\varepsilon_g(\Omega_f) = I(z_g) \tag{76}$$

where $b(\Omega_f) \equiv b_{\mathbf{u}_z}(\Omega_f)$, like $\varepsilon(\Omega_f,\mathbf{r})$, pertains to a horizontal leaf (with normal $\Omega_L = \mathbf{u}_z$):

$$b(\Omega_f) \equiv \hat{\mu}_f^{-1} \int d\Omega_i \, \hat{\mu}_i \, \sigma(\Omega_i \to \Omega_f) \tag{77}$$

Thus, a canopy of leaves and ground obeying (76) does not disturb an isotropic climate $I(\Omega,\mathbf{r}) = I(\mathbf{r})$. Moreover, it does not cause $I(\mathbf{r})$ to vary with $\mathbf{r}$, since the right side of (73) vanishes if (76) holds, so that $\nabla I(\mathbf{r}) = 0$. It follows that if, for all $\Omega_f$ and all $\mathbf{r}$:

$$\varepsilon(\Omega_f,\mathbf{r}) = \hat{\mu}_f\big(1 - b(\Omega_f)\big)I_0 \quad \text{and} \quad \varepsilon_g(\Omega_f) = \hat{\mu}_f\big(1 - b_g(\Omega_f)\big)I_0 \tag{78}$$

then the uniform isotropic radiance $I(\Omega,\mathbf{r}) = I_0$ satisfies the LTE, and is the solution *provided* it is compatible with boundary conditions. In the horizontally homogeneous case, this requires the incident light at $z_0$ to be 'down' isotropic, $D(\Omega,z_0) = I_0$.

With leaves *not* satisfying (76), it is still possible to have an isotropic radiance, provided $\eta(\Omega_L,\mathbf{r})$ is such that the sum over leaf orientations averages out the disturbances caused by individual leaves (so that *collectively* the leaves do not disrupt an isotropic radiance). Isotropic leaf orientations immediately come to mind.

***Isotropic leaf orientations***: With $\eta(\Omega_L,\mathbf{r}) = \frac{1}{4\pi}\eta(\mathbf{r})$, the local canopy scattering function $S(\Omega_i \to \Omega_f,\mathbf{r})$ is a function of only the angle between $\Omega_i$ and $\Omega_f$, so that its integral over $\Omega_i$ is equal to that over $\Omega_f$, hence to $\Gamma(\mathbf{r}) - A(\mathbf{r})$, by (19). Then equation (73) becomes

$$\Omega \cdot \nabla I(\mathbf{r}) = \rho(\Omega,\mathbf{r}), \qquad \rho(\Omega,\mathbf{r}) \equiv \mathcal{E}(\Omega,\mathbf{r}) - A(\mathbf{r})I(\mathbf{r}) \tag{79}$$

where $\rho(\Omega,\mathbf{r})$ is the net gain (emission minus absorption) of photons $\Omega$ at $\mathbf{r}$. Now, if $I(\Omega,\mathbf{r}) = I(\mathbf{r})$ is isotropic, so will the thermal emission $\mathcal{E}(\Omega,\mathbf{r}) = \frac{1}{4\pi}\mathcal{E}(\mathbf{r})$ (since the leaves are isotropically oriented), whence also $\rho(\Omega, \mathbf{r}) \equiv \rho(\mathbf{r})$. Then equation (79) makes sense only if $\rho(\mathbf{r}) \equiv 0$, for it says that if, e.g., $\rho(\mathbf{r}) > 0$, then $I(\mathbf{r})$ increases in *all* directions (hence in both $\Omega$ and $-\Omega$, for any $\Omega$). Thus, with isotropic leaf orientations, if $\mathcal{E}(\Omega, \mathbf{r}) = A(\mathbf{r}) I_0$, then an isotropic radiance $I(\Omega, \mathbf{r}) = I_0$ *uniform* in $\mathbf{r}$ satisfies the LTE, and is the solution under the proper boundary conditions, whatever the form of $\sigma(\Omega_i \to \Omega_f)$.

***Lambertian leaves and ground***: Let us now return to the individual leaf requirement (76). This is compatible with only certain kinds of leaf optical coefficients, such as Lambertian (specular will also do). By inserting the Lambertian coefficients (57),(59) into (75) we get:[2]

$$b(\Omega_f) = b_{\mathrm{sgn}\,\mu_f}, \qquad b_+ \equiv \hat{\sigma}_{++} + \hat{\sigma}_{-+}, \qquad b_- \equiv \hat{\sigma}_{--} + \hat{\sigma}_{+-}$$
$$b_g(\Omega_f) = \hat{\sigma}_g \tag{80}$$

so that (76) may be written, refering to (54):

$$\tfrac{1}{\pi}\hat{\varepsilon}_\pm(\mathbf{r}) = (1 - b_\pm) I(\mathbf{r}), \qquad \hat{\varepsilon}_g = (1 - \hat{\sigma}_g) I(z_g) = \hat{\alpha}_g I(z_g) \tag{81}$$

Noting now that the integral of (60) over $\Omega_i$ may be written (compare (61))

$$\int d\Omega_i\, S(\Omega_i \to \Omega_f, \mathbf{r}) = b_+ \Gamma_+(\Omega_f, \mathbf{r}) + b_- \Gamma_-(\Omega_f, \mathbf{r}) \tag{82}$$

and using (65), we may rewrite (73) as

$$\Omega_f \cdot \nabla I(\mathbf{r}) = \Gamma_+(\Omega_f, \mathbf{r})\left[\tfrac{1}{\pi}\hat{\varepsilon}_+(\mathbf{r}) - (1 - b_+) I(\mathbf{r})\right] + \Gamma_-(\Omega_f, \mathbf{r})\left[\tfrac{1}{\pi}\hat{\varepsilon}_-(\mathbf{r}) + (1 - b_-) I(\mathbf{r})\right] \tag{83}$$

If (81) holds, then again the right side of (83) vanishes, so that $\nabla I(\mathbf{r}) = 0$, and $I(\Omega, \mathbf{r}) = I_0$ is the solution under proper boundary conditions, whatever the distribution of leaf orientations. By restricting the latter, we will next obtain isotropic radiances that are *non-uniform* in *z*.

***$\varphi_L$-symmetric Lambertian canopy***: Using $\Gamma_\pm = \tfrac{1}{2}\Gamma \pm \tfrac{1}{2}\Delta\Gamma$ and $b_+ + b_- = 2 - \hat{\alpha}$ together with (56), rewrite (83) in the horizontally homogeneous case as

$$\mu_f \frac{d}{dz} I(z) = \tfrac{1}{2} \Gamma(\Omega_f, z)\left[\tfrac{1}{\pi}\widehat{\varepsilon}(z) - \widehat{\alpha} I(z)\right] + \tfrac{1}{2} \Delta\Gamma(\Omega_f, z)\left[\tfrac{1}{\pi}\Delta\widehat{\varepsilon}(z) + (b_+ - b_-)I(z)\right] \quad (84)$$

---

[2] Note that $b_+ = \hat{\sigma}_{++} + \hat{\sigma}_{-+}$ pertains to what the leaf bottom '*emits*', since $\hat{\sigma}_{++}$ is the fraction of 'down' light transmitted through from the other side, and $\hat{\sigma}_{-+}$ the fraction of incident 'up' light reflected back down. By contrast, $\hat{\sigma}_{++} + \hat{\sigma}_{+-} = 1 - \hat{\alpha}_+$ is the fraction of light *incident* on the top of the leaf that gets scattered (hence is not absorbed). If there is zero emission, then (81) implies $b_+ = b_- = 1$, hence $b_+ + b_- = 2 - \hat{\alpha}_+ - \hat{\alpha}_- = 2$, hence $\hat{\alpha}_\pm = 0$ (since $\hat{\alpha}_\pm \leq 1$), i.e., zero absorption. Note that $\hat{\alpha}_\pm = 0$ do *not* imply $b_+ = b_- = 1$: For instance, $\hat{\sigma}_{++} = \tfrac{3}{4}$, $\hat{\sigma}_{+-} = \tfrac{1}{4}$, $\hat{\sigma}_{--} = \tfrac{1}{4}$, $\hat{\sigma}_{-+} = \tfrac{3}{4}$ yield $\hat{\alpha}_\pm = 0$, but $b_+ = \tfrac{3}{2}$, $b_- = \tfrac{1}{2}$.

Assume now that $\eta(\Omega_L, z) = \tfrac{1}{2\pi}\tilde{\eta}(\mu_L, z)$ is $\varphi_L$-symmetric. Then, $\Delta\Gamma(\Omega_f, z) = \eta(z)\mu_f \langle \mu_L \rangle_z$ by (31), where $\langle \mu_L \rangle_z$ is the mean leaf inclination at height $z$. It follows that if $\tfrac{1}{\pi}\widehat{\varepsilon}(z) - \widehat{\alpha} I(z) = 0$, then (84) can be divided by $\mu_f$, and becomes:

$$\frac{d}{dz} I(z) = \tfrac{1}{2} \eta(z)\langle \mu_L \rangle_z \left[\tfrac{1}{\pi}\Delta\widehat{\varepsilon}(z) + (b_+ - b_-)I(z)\right] \quad \text{if} \quad \tfrac{1}{\pi}\widehat{\varepsilon}(z) = \widehat{\alpha} I(z) \quad (85)$$

Let us now (artificially) assume emissions proportional to $I(z)$, by setting

$$\widehat{\varepsilon}_\pm(z) = \pi c_\pm I(z), \qquad \widehat{\varepsilon}_g = \pi c_g I(z_g), \qquad c \equiv c_+ + c_- \quad (86)$$

The condition $\tfrac{1}{\pi}\widehat{\varepsilon}(z) = \widehat{\alpha} I(z)$ then becomes $c = \widehat{\alpha}$, and we get, putting $\bar{b}_\pm \equiv b_\pm + c_\pm$:

$$\frac{d}{dz} I(z) = \tfrac{1}{2} \eta(z)\langle \mu_L \rangle_z (\bar{b}_+ - \bar{b}_-) I(z) \quad \text{if} \quad c = \widehat{\alpha} \Leftrightarrow \bar{b}_+ + \bar{b}_- = 2 \quad (87)$$

$$I(\Omega, z) = I(z) = I(z_0) e^{\tfrac{1}{2}\int_{z_0}^{z} dz'\, \eta(z')\langle \mu_L \rangle_{z'} (\bar{b}_+ - \bar{b}_-)}, \qquad \bar{b}_\pm \equiv b_\pm + c_\pm \quad (88)$$

The radiance (88) is the solution of (87) provided boundary conditions are compatible with it, namely $D(\Omega, z_0) \equiv I_0$ is 'down' isotropic, and $\widehat{\varepsilon}_g = \widehat{\alpha}_g I(z_g)$ in (81) holds, i.e., $c_g = \widehat{\alpha}_g$. In numerical tests, we will substitute (88) into (86), and verify how accurately our numerical integration of the LTE (40),(44), with *these* emissions, gives back the light climate (88). Note that if the distribution $\tilde{\lambda}(\mu_L, z) = \tilde{\lambda}(\mu_L)$ of leaf inclinations is uniform in $z$, then

$$I(z) = I(z_0) e^{\tfrac{1}{2}\mathcal{L}(z)\langle \mu_L \rangle (\bar{b}_+ - \bar{b}_-)}, \qquad \mathcal{L}(z) \equiv \mathcal{L}(z_0, z) \quad (89)$$

where $\mathcal{L}(z)$ is the LAI above $z$. If $\tilde{\lambda}(\mu_L, z) = \tfrac{1}{2}$ is isotropic, so that $\langle \mu_L \rangle_z = 0$, or if $\bar{b}_+ = \bar{b}_- = 1$, implying (81), then $I(z) = I(z_0) = I_0$ is uniform, as found after Eq.(79).[3]

*Example*: *One dimensional model*: Suppose all photons travel along the $z$ axis, so that $\Omega = \pm \mathbf{u}_z$. Assume also that $\eta(z) = 1$ and $\langle \mu_L \rangle_z = 1$ for all $z$, and that all $\widehat{\sigma}$'s are zero, i.e., there is no scattering, so that $\hat{\alpha}_+ = \hat{\alpha}_- = \hat{\alpha}_g = 1$. Then the 'down' and 'up' radiances at $D(z)$ and $U(z)$ at $z$ consist of the incident radiance $D_0$ at $z = z_0$, plus light emitted above or below $z$, attenuated due to interception by leaves, that is:

$$D(z) = D_0 e^{-(z-z_0)} + \int_{z_0}^{z} dz'\, \mathcal{E}_D(z') e^{-(z-z')}, \qquad U(z) = \varepsilon_g e^{z-z_g} + \int_{z}^{z_g} dz'\, \mathcal{E}_U(z') e^{z-z'} \qquad (90)$$

where $\mathcal{E}_D(z) = \mathcal{E}(\mathbf{u}_z, z)$ and $\mathcal{E}_U(z) = \mathcal{E}(-\mathbf{u}_z, z)$ are the 'down' and 'up' components of the local canopy emission $\mathcal{E}(\Omega, z)$. If the isotropic radiance conditions are satisfied, i.e., if $c = \hat{\alpha} = 2$ and $c_g = \hat{\alpha}_g = 1$, see (87), then (88) becomes $I(z) = D_0 e^{\frac{1}{2}(c_+ - c_-)(z - z_0)}$. By then plugging $\mathcal{E}_D(z) = c_+ I(z)$, $\mathcal{E}_U(z) = c_- I(z)$, and $\varepsilon_g = I(z_g)$ into (90), and doing the integrals using $c_+ - c_- \pm 2 = \pm 2 c_\pm$, one indeed gets $D(z) = U(z) = I(z)$.

---

[3] Note that $\bar{b}_+ = \bar{b}_- = 1$ implies $c_\pm = \widehat{\alpha}_\pm \pm (\widehat{\sigma}_{-+} - \widehat{\sigma}_{+-})$, hence $c_\pm = \widehat{\alpha}_\pm$ if $\widehat{\sigma}_{-+} = \widehat{\sigma}_{+-}$, i.e., emission by each side of the leaf equals absorption. With $\widehat{\alpha}_\pm = 1$, this corresponds to black body leaves at a given temperature immersed in thermal light of the same temperature.

## 11 Canopy of horizontal Lambertian leaves

In this and the next section, we will refer to equations in Ref.[5], which will be written as I-(10), for instance. We here use the notations $\hat{\sigma}_{++} = \hat{\sigma}_{dd}$, $\hat{\sigma}_{+-} = \hat{\sigma}_{du}$, etc.

With horizontal Lambertian leaves, the canopy optical coefficients are given by

$$\Gamma(\Omega, z) = \eta(z)\hat{\mu}, \qquad \mathcal{E}(\Omega_f, z) = \pi^{-1} \eta(z) \hat{\mu}_f \hat{\varepsilon}_{\mathrm{sgn}\,\mu_f}$$
$$S(\Omega_i \to \Omega_f, z) = \pi^{-1} \eta(z) \hat{\mu}_i \hat{\mu}_f \hat{\sigma}_{\mathrm{sgn}\,\mu_f, \mathrm{sgn}\,\mu_i} \qquad (91)$$

in view of (32) and (69). So the LTE (40) divided by $\hat{\mu}_f$, and (44), become:

$$\mathrm{sgn}\,\mu_f \frac{d}{dz} I(\Omega_f, z) = \eta(z)\left[-I(\Omega_f, z) + \tfrac{1}{\pi} \int d\Omega_i\, I(\Omega_i, z) \hat{\mu}_i \hat{\sigma}_{\mathrm{sgn}\,\mu_i, \mathrm{sgn}\,\mu_f} + \tfrac{1}{\pi} \hat{\varepsilon}_{\mathrm{sgn}\,\mu_f}(z)\right]$$
$$U(\Omega_f, z_g) = \tfrac{1}{\pi} \int_{\mu_i > 0} d\Omega_i\, D(\Omega_i, z_g) \hat{\mu}_i \hat{\sigma}_g + \tfrac{1}{\pi} \hat{\varepsilon}_g = \tfrac{1}{\pi}\left(\hat{\sigma}_g \mathcal{D}_g + \hat{\varepsilon}_g\right) \qquad (92)$$

This will now be solved analytically, in the absence of emissions.

***LTE for the total down and up fluxes***: Multiplying (92) by $\hat{\mu}_f$, and integrating over $\Omega_f \in D$ or $\Omega_f \in U$, we get equations for the total vertical fluxes $\mathcal{D}(z)$ and $\mathcal{U}(z)$:

$$\frac{d}{dz}\begin{pmatrix} \mathcal{D}(z) \\ -\mathcal{U}(z) \end{pmatrix} = \eta(z) \mathbf{M} \begin{pmatrix} \mathcal{D}(z) \\ \mathcal{U}(z) \end{pmatrix} + \begin{pmatrix} \mathcal{E}_D(z) \\ \mathcal{E}_U(z) \end{pmatrix}, \qquad \mathbf{M} \equiv \begin{pmatrix} -1 + \hat{\sigma}_{dd} & \hat{\sigma}_{ud} \\ \hat{\sigma}_{du} & -1 + \hat{\sigma}_{uu} \end{pmatrix}$$
$$\mathcal{U}_g = \hat{\sigma}_g \mathcal{D}_g + \hat{\varepsilon}_g \qquad (93)$$

(we used (4) and (39)) where the 'down' and 'up' emitted vertical fluxes are

$$\mathcal{E}_D(z) = \eta(z)\widehat{\varepsilon}_+(z), \qquad \mathcal{E}_U(z) = \eta(z)\widehat{\varepsilon}_-(z) \tag{94}$$

Eq.(93) has the form of the matrix transport equation I-(11) in Ref.[5], but in two dimensions (up and down). The transfer matrix for a layer of LAI $= \mathcal{L}$ is, on using (A.4) and $\mathbf{s} \equiv \begin{pmatrix} 1 & 0 \\ 0 & -1 \end{pmatrix}$:

$$\mathbf{T} = e^{\mathcal{L}\mathbf{sM}} = e^{\frac{1}{2}\mathcal{L}Tr(\mathbf{sM})}\left[\mathbf{1}\cosh\lambda + \mathcal{L}\overline{\mathbf{M}}(\sinh\lambda)/\lambda\right], \qquad Tr(\mathbf{sM}) = \widehat{\sigma}_{dd} - \widehat{\sigma}_{uu}$$

$$\overline{\mathbf{M}} \equiv \mathbf{sM} - \tfrac{1}{2}\mathbf{1}Tr(\mathbf{sM}) = \begin{pmatrix} -1 + \tfrac{1}{2}(\widehat{\sigma}_{dd} + \widehat{\sigma}_{uu}) & \widehat{\sigma}_{ud} \\ -\widehat{\sigma}_{du} & 1 - \tfrac{1}{2}(\widehat{\sigma}_{dd} + \widehat{\sigma}_{uu}) \end{pmatrix} \tag{95}$$

$$\lambda \equiv \sqrt{-\det\mathcal{L}\overline{\mathbf{M}}} = \mathcal{L}\sqrt{\Delta}, \qquad \Delta \equiv -\det\overline{\mathbf{M}} = \left(1 - \tfrac{1}{2}\widehat{\sigma}_{dd} - \tfrac{1}{2}\widehat{\sigma}_{uu}\right)^2 - \widehat{\sigma}_{du}\widehat{\sigma}_{ud} \geq 0$$

Note that $\Delta \geq \left(1 - \tfrac{1}{2}\widehat{\sigma}_{dd} - \tfrac{1}{2}\widehat{\sigma}_{uu}\right)^2 - (1 - \widehat{\sigma}_{dd})(1 - \widehat{\sigma}_{uu}) = \tfrac{1}{4}(\widehat{\sigma}_{dd} - \widehat{\sigma}_{uu})^2 \geq 0$ since $\widehat{\sigma}_{du} \leq 1 - \widehat{\sigma}_{dd}$ and $\widehat{\sigma}_{ud} \leq 1 - \widehat{\sigma}_{uu}$, by (55), so that $\lambda$ is always real. If there is zero emission, then the flux vector $\mathbf{J}(z) = (\mathcal{D}(z), \mathcal{U}(z))$ is given by $\mathbf{J}(z) = \mathbf{T}(z, z_0)\mathbf{J}(z_0)$, see I-(14), and the 'up' flux $\mathcal{U}(z_0)$ can be obtained analytically by using I-(19,20), since here $T_{DD}$, $T_{DU}$, etc., are scalars.

The vertical fluxes $\mathcal{D}(z)$ and $\mathcal{U}(z)$ so obtained in fact provide the complete solution. Indeed, because each horizontal layer $dz$ of canopy acts as a Lambertian surface, all *scattered* (and emitted if any) radiance is semi-isotropic, i.e., of the form

$$D^{scatt}(\Omega, z) = \tfrac{1}{\pi}\mathcal{D}^{scatt}(z), \qquad U^{scatt}(\Omega, z) = \tfrac{1}{\pi}\mathcal{U}^{scatt}(z) \tag{96}$$

The only radiance possibly *not* semi-isotropic is the *unscattered* light

$$D^{(0)}(\Omega, z) \equiv e^{-\mathcal{L}(z)}D(\Omega, z_0), \qquad \mathcal{L}(z) \equiv \mathcal{L}(z_0, z) \tag{97}$$

where we recalled, see after Eq.(37), that in the case of horizontal leaves:

$$\xi_\Omega(z_1, z_2) = e^{-\mathcal{L}(z_1, z_2)}, \qquad \mathcal{L}(z_1, z_2) = \left|\int_{z_1}^{z_2} dz\, \eta(z)\right| \tag{98}$$

Now, $\mathcal{U}^{scatt}(z) = \mathcal{U}(z)$ (all 'up' light is necessarily scattered), while $\mathcal{D}^{scatt}(z) = \mathcal{D}(z) - e^{-\mathcal{L}(z)}\mathcal{D}_0$, where $\mathcal{D}_0 = \mathcal{D}(z_0)$ is the vertical flux at $z_0$. So finally:

$$U(\Omega, z) = \tfrac{1}{\pi}\mathcal{U}(z), \qquad D(\Omega, z) = e^{-\mathcal{L}(z)}D(\Omega, z_0) + \tfrac{1}{\pi}\left(\mathcal{D}(z) - e^{-\mathcal{L}(z)}\mathcal{D}_0\right) \tag{99}$$

If $D(\Omega, z_0) = \tfrac{1}{\pi}\mathcal{D}_0$ is 'down' isotropic, see (48), then $D(\Omega, z) = \tfrac{1}{\pi}\mathcal{D}(z)$. Note that substituting $I(\Omega, z) = \{\tfrac{1}{\pi}\mathcal{D}(z), \tfrac{1}{\pi}\mathcal{U}(z)\}$ into (92) gives back Eq.(93). Note also that if the leaves and ground are non-emitting and non-absorbing, then $\mathcal{U}(z) = \mathcal{D}(z)$ by (46), so that $I(\Omega, z) = \tfrac{1}{\pi}\mathcal{D}(z)$ is fully isotropic, provided $D(\Omega, z_0) = \tfrac{1}{\pi}\mathcal{D}_0$ is 'down' isotropic.

## 12 A case of extreme light trapping

Consider the following situation:

(i) There are no intrinsic emissions by the leaves or by the ground.

(ii) All leaves are horizontal, completely transparent from the top, $\hat{\sigma}_{dd} = 1$, and totally reflective from the bottom, $\hat{\sigma}_{ud} = 1$. Thus, $\hat{\sigma}_{du} = \hat{\sigma}_{uu} = 0$, and there is zero absorption.

(iii) The ground is perfectly reflective: $R_g \equiv \int_{\mu_f > 0} d\Omega_f \sigma_g(\Omega_i \to \Omega_f) = 1$.

This artificial situation will be treated in some detail, as a two-dimensional illustration of the methods discussed in Ref.[5]. We let $\mathcal{L}(z) \equiv \mathcal{L}(z_0, z)$ be the LAI above $z$, and $\mathcal{L} \equiv \mathcal{L}(z_0, z_g)$ be the total LAI of the canopy. We will need the formula

$$\sum_{n=1}^{N} a^{N-n} = \sum_{m=0}^{N-1} a^m = (1-a^N)/(1-a) \tag{100}$$

***Heuristics***: Since $\hat{\sigma}_{dd} = 1$, all 'down' light $\mathcal{D}_0 = \mathcal{D}_0^d + \mathcal{D}_0^h$ incident on the top $z_0$ of the canopy reaches the ground $z_g$, where it is totally reflected. A fraction $\tau = e^{-\mathcal{L}}$ of that reflected light makes it to $z_0$ unimpeded, and escapes. The rest hits some leaf, and is reflected back down to $z_g$, where it is reflected back up, and so on. Since there is no absorption and no emission, what comes out must equal what comes in, $\mathcal{U}_0 = \mathcal{D}_0$, hence $\mathcal{U}(z) = \mathcal{D}(z)$ at all $z$, by (41).

This is similar to a greenhouse with a perfectly reflecting surface at $z_g$, and at $z_0$ a membrane letting all 'down' light from above through, but reflecting all 'up' light from below, except for a fraction $\tau$ that is let through. So radiation builds up between the two surfaces, until it is sufficiently intense that what escapes equals what comes in, i.e., until $\mathcal{U}_0 = \tau \mathcal{U}_g = \mathcal{D}_0$, or $\mathcal{U}_g = e^{\mathcal{L}} \mathcal{D}_0$. For small $\tau = e^{-\mathcal{L}}$, this buildup requires many back and forth reflections. Since the $k$th reflection adds $(1-\tau)^k \mathcal{D}_0$ to $\mathcal{U}_g$, we indeed get, putting $r \equiv 1 - \tau$:

$$\mathcal{U}_g = (1 + r + r^2 + ...)\mathcal{D}_0 = (1-r)^{-1}\mathcal{D}_0 = \tau^{-1}\mathcal{D}_0, \qquad r \equiv 1-\tau, \quad \tau \equiv e^{-\mathcal{L}} \tag{101}$$

Let $k$ be the number of terms in the series (101) (hence of reflections) required to approximate $\tau^{-1}\mathcal{D}_0$ to within some relative error $\delta$. Since

$$s \equiv \tau^{-1} = s_k + \sum_{n=k+1}^{\infty} r^n = s_k + r^{k+1} s \qquad \text{where} \qquad s_k \equiv \sum_{m=0}^{k} r^m \tag{102}$$

the relative error is $\delta \equiv (s-s_k)/s = r^{k+1} = (1-\tau)^{k+1}$, so that we require

$$k+1 \geq (\ln \delta)/\ln(1-\tau) \approx -\tau^{-1}(\ln \delta) = -e^{\mathcal{L}}(\ln \delta) \tag{103}$$

since $\ln(1-\tau) \approx -\tau$ for small $\tau$. For instance, $\mathcal{L} = 10$ and $\delta = 10^{-3}$ yield $e^{\mathcal{L}} \approx 22\,000$ and $k \approx 152\,000$. Thus, for instance, Monte Carlo computations would be prohibitively long, since each photon will rebound that many times, on average, before escaping.

***Remark***: Since fluxes get attenuated along their direction of propagation, radiances usually decrease from the top $z_0$ to the bottom $z_g$ of the canopy. But in the present situation, the radiances *increase* from $z_0$ to $z_g$, because light is not attenuated on the way down, while all 'up' light hitting a leaf gets reflected down to the ground.

***Explicit solutions***: We next solve the LTE (93) by using the three main methods of Ref.[5], namely the direct transfer matrix method, the Green's matrix method, and iterative integration. In order to illustrate these methods with emission terms present, we will treat first-scattered sunlight as an 'emission'. Since $\mathcal{D}^h(z) = e^{-\mathcal{L}(z)}\mathcal{D}_0^h$, and since any sunlight intercepted by a leaf just goes straight through it, hence is 're-emitted' downwards, the local canopy 'sun emissions', integrated either over $\Omega_f \in D$ or over $\Omega_f \in U$, are:

$$\mathcal{E}_D^h(z) = e^{-\mathcal{L}(z)}\mathcal{D}_0^h \eta(z), \qquad \mathcal{E}_U^h(z) = 0, \qquad e_g^h = e^{-\mathcal{L}}\mathcal{D}_0^h \tag{104}$$

***Transfer matrix method***: Here, $\mathbf{M} = \begin{pmatrix} 0 & 1 \\ 0 & -1 \end{pmatrix}$ in (93), so that $\overline{\mathbf{M}} = \begin{pmatrix} -1/2 & 1 \\ 0 & 1/2 \end{pmatrix}$ and $\Delta = \tfrac{1}{4}$ in (95). Thus, by (95) and I-(14)(b), the transfer matrix and propagated emissions are

$$\mathbf{T}(z,z_0) = \begin{pmatrix} T_{DD} & T_{DU} \\ T_{UD} & T_{UU} \end{pmatrix} = \begin{pmatrix} 1 & e^{\mathcal{L}(z)}-1 \\ 0 & e^{\mathcal{L}(z)} \end{pmatrix}, \qquad \mathbf{f}(z) = \begin{pmatrix} f_D \\ f_U \end{pmatrix} = \begin{pmatrix} (1-e^{-\mathcal{L}(z)})\mathcal{D}_0^h \\ 0 \end{pmatrix} \tag{105}$$

where $f_D(z) = \mathcal{D}_0^h \int_{z_0}^z dz' \eta(z') e^{-\mathcal{L}(z')} = \mathcal{D}_0^h(1-e^{-\mathcal{L}(z)})$ and $f_U(z) = 0$, by (98) and since $\mathbf{E}^h(z) = (\mathcal{E}_D^h(z), 0)$, so that $\mathbf{T}(z,z')\mathbf{E}^h(z') = (\mathcal{E}_D^h(z'), 0)$. With $R_g = 1$ and $e_g = e_g^h = e^{-\mathcal{L}}\mathcal{D}_0^h$ in I-(20), we get (here, $R_g$, $T_{DD}$, $T_{DU}$, etc., are scalars):

$$R_0 = \frac{R_g T_{DD} - T_{UD}}{T_{UU} - R_g T_{DU}} = \frac{1}{1} = 1, \qquad e_0 = \frac{R_g f_D - f_U + e_g}{T_{UU} - R_g T_{DU}} = \mathcal{D}_0^h \tag{106}$$

whence $\mathcal{U}_0 = \mathcal{U}_0^d = R_0 \mathcal{D}_0^d + e_0 = \mathcal{D}_0^d + \mathcal{D}_0^h = \mathcal{D}_0$, as should be. Also, by I-(14)(a):

$$\begin{pmatrix} \mathcal{D}^d(z) \\ \mathcal{U}(z) \end{pmatrix} = \mathbf{T}(z,z_0) \begin{pmatrix} \mathcal{D}_0^d \\ \mathcal{U}_0 \end{pmatrix} + \mathbf{f}(z) = \begin{pmatrix} e^{\mathcal{L}(z)}\mathcal{D}_0 - \mathcal{D}^h(z) \\ e^{\mathcal{L}(z)}\mathcal{D}_0 \end{pmatrix}, \qquad \mathcal{D}_0 = \mathcal{D}_0^d + \mathcal{D}_0^h \qquad (107)$$

where $\mathcal{D}^h(z) = e^{-\mathcal{L}(z)}\mathcal{D}_0^h$. Thus, $\mathcal{D}(z) = \mathcal{D}^d(z) + \mathcal{D}^h(z) = e^{\mathcal{L}(z)}\mathcal{D}_0$, that is, radiance increases exponentially *down* the canopy (light trapping). Note that if $\mathcal{L}$ is large, then the denominator in (106), namely $T_{UU} - R_g T_{DU} = e^{\mathcal{L}} - (e^{\mathcal{L}} - 1) = 1$, is the difference of two *large* numbers, $e^{\mathcal{L}}$ and $e^{\mathcal{L}} - 1$. So numerical nonsense is expected whenever $e^{\mathcal{L}} > 10^{14}$ if 15 digits are carried in numerical computations (in double precision), i.e., if $\mathcal{L} > 14 \ln 10 \approx 30$.

*Green's matrix method*: The transmission-reflection matrices and emission vectors for the whole canopy follow from I-(22) and (105) with $z = z_g$ (they are rather obvious):

$$\begin{pmatrix} t & r \\ \rho & \tau \end{pmatrix} = \begin{pmatrix} 1 & 1-\tau \\ 0 & \tau \end{pmatrix}, \qquad \begin{pmatrix} d \\ u \end{pmatrix} = \begin{pmatrix} (1-\tau)\mathcal{D}_0^h \\ 0 \end{pmatrix}, \qquad \tau = e^{-\mathcal{L}} \qquad (108)$$

where $\tau$ is the 'up' transmission through the whole canopy. Note that the total 'emission' $d$ from the bottom of the canopy (due to first-scattered sunlight 'emission') is equal to the *intercepted* sunlight $(1-\tau)\mathcal{D}_0^h$, because any sunlight intercepted by a leaf goes straight through it, hence is 're-emitted' downwards, and is then propagated without attenuation down to $z_g$ (that is, out of the foliage). Eqs. I-(33), in the case of a single 'medium' layer, become:

$$\mathcal{E} = \begin{pmatrix} t\mathcal{D}_0^d + d \\ e_g \end{pmatrix} = \begin{pmatrix} \mathcal{D}_0^d + (1-\tau)\mathcal{D}_0^h \\ \tau \mathcal{D}_0^h \end{pmatrix}, \qquad \mathcal{Q} = \begin{pmatrix} 0 & r \\ R_g & 0 \end{pmatrix} = \begin{pmatrix} 0 & 1-\tau \\ 1 & 0 \end{pmatrix} \qquad (109)$$

Noting that $\mathcal{Q}^2 = (1-\tau)\mathbf{1}$, hence $(1+\mathcal{Q})(1-\mathcal{Q}) = 1 - \mathcal{Q}^2 = \tau\mathbf{1}$, we deduce that $\mathcal{G} \equiv (1-\mathcal{Q})^{-1} = \tau^{-1}(1+\mathcal{Q})$. Here, $\mathcal{J} = (\mathcal{D}_g^d, \mathcal{U}_g)$, see I-(33), and $\mathcal{J} = \mathcal{G}\mathcal{E}$ in I-(34) yields (107) with $z = z_g$. Note that $\mathcal{G}$ is not obtainable numerically if 15 digits are carried and $\mathcal{L} > 30$, for then $\tau < 10^{-14}$ and $1-\tau$ is indistinguishable from 1. But of course, we would then use several 'medium' layers (indexed by $m$), such that no $r_m = 1 - \tau_m$ be too close to 1. To have $e^{\mathcal{L}_m} < 10^7$ say, these medium layers should be of LAI $\mathcal{L}_m < 7 \ln 10 \approx 15$.

*Iterative integration*: With $N$ thin layers of $LAI = \mathcal{L}/N$, the thin-layer transmission-reflection matrices are given by (108) with $\tau$ replaced by $\Delta\tau = e^{-\mathcal{L}/N}$. Since the attenuated sunlight incident on the $n$th layer (counting from the top $z_0$) is $(\Delta\tau)^{n-1}\mathcal{D}_0^h$, and $(\Delta\tau)^N = \tau$, the iteration equations I-(39) read in the present case:

$$\mathcal{D}_n^d = \mathcal{D}_{n-1}^d + (1-\Delta\tau)\mathcal{U}_n + (1-\Delta\tau)(\Delta\tau)^{n-1}\mathcal{D}_0^h$$
$$\mathcal{U}_N = \mathcal{D}_N^d + \tau\mathcal{D}_0^h \qquad (\Delta\tau \equiv e^{-\mathcal{L}/N}) \qquad (110)$$
$$\mathcal{U}_{n-1} = \Delta\tau\,\mathcal{U}_n$$

where $\mathcal{D}_N = \mathcal{D}_g \equiv \mathcal{D}(z_g)$ and $\mathcal{U}_N = \mathcal{U}_g \equiv \mathcal{U}(z_g)$. It follows that

$$\mathcal{U}_n = (\Delta\tau)^{N-n}\mathcal{U}_N, \qquad \mathcal{D}_n^d = \mathcal{D}_0^d + (1-\Delta\tau)\sum_{m=1}^n \left(\mathcal{U}_m + (\Delta\tau)^{m-1}\mathcal{D}_0^h\right) \qquad (111)$$

expressing that $\mathcal{U}$ decreases from $z_g$ to $z_0$, since any 'up' light hitting a leaf gets reflected down to $z_g$, while 'down' light increases from $z_0$ to $z_g$ since nothing blocks it, and it picks up on the way down all downwards reflected 'up' light, as well as all first-scattered sunlight 'emissions'. Doing the sums in (111) (with $n = N$), by using (100) and $\tau \equiv (\Delta\tau)^N$, we get:

$$\mathcal{D}_N^d - \mathcal{D}_0^d = (1-\Delta\tau)\sum_{n=1}^N \left((\Delta\tau)^{N-n}\mathcal{U}_N + (\Delta\tau)^{n-1}\mathcal{D}_0^h\right) = (1-\tau)\left(\mathcal{U}_N + \mathcal{D}_0^h\right) \qquad (112)$$

Using $\mathcal{D}_N^d = \mathcal{U}_N - \tau\mathcal{D}_0^h$ from (110), and $\mathcal{D}_0^d = \mathcal{D}_0 - \mathcal{D}_0^h$, we rewrite (112) as

$$\mathcal{U}_g = \mathcal{D}_0 + r\,\mathcal{U}_g \qquad \text{where} \qquad r = 1-\tau \qquad (113)$$

Thus, if $\mathcal{U}_g^{(k)}$ is the value of $\mathcal{U}_g$ obtained after the $k$th iteration, then $\mathcal{U}_g^{(k)} = \mathcal{D}_0 + r\,\mathcal{U}_g^{(k-1)}$. So with $\mathcal{U}_g^{(0)} \equiv 0$, successive iterations yield

$$\mathcal{U}_g^{(1)} = \mathcal{D}_0, \qquad \mathcal{U}_g^{(2)} = \mathcal{D}_0 + r\mathcal{D}_0, \qquad \cdots \qquad \mathcal{U}_g^{(k)} = \mathcal{D}_0 \sum_{m=0}^k r^m \qquad (114)$$

so that again, $\mathcal{U}_g^{(k)} \to \tau^{-1}\mathcal{D}_0$ as $k \to \infty$. The Green's matrix method amounts to solving (113) directly, namely $\mathcal{U}_g = (1-r)^{-1}\mathcal{D}_0 = \tau^{-1}\mathcal{D}_0$. Note that rewriting (114) as $\mathcal{U}_g^{(k)} = \mathcal{U}_g^{(k-1)} + r^k\mathcal{D}_0$ shows that the $k$th iteration adds in $k$th scattered light.

In the above, we used the exact thin-layer transmission-reflection matrices, namely (108) with $\Delta\tau$ instead of $\tau$. If these were treated to first order in the thin-layer thickness $\Delta z$, as usually has to be done in realistic numerical computations, then as argued after Eq. I-(46), the thin-layer transmissions would be too small, and the reflections too large. This would lead to too much light trapping in the present situation.

## 13  Conclusion

The numerical integration methods proposed in Ref.[5] will be used in subsequent papers to compute light climates and rates of photosynthesis in realistic canopies. But the accuracy of our methods will first be tested by applying them to the models discussed in the present paper, and then comparing the numerical results with the analytical solutions. We will find that accuracy, as well as speed, are generally much better than provided by the widely used iterative integration [1-3]. The latter may in fact become impractical in certain situations, such as extreme light trapping, while our method remains as efficient. For *realistic* canopies, however, iterative integration is of sufficient accuracy (except perhaps in the case of grazing sunlight). But the fact that our method is faster (as well as more accurate) is a significant advantage when repeated integrations of the LTE must be done, as when iterating over leaf temperatures.

Let us end by indicating ways in which the basic theory (in Part I) can be generalized:

(i) To allow for several kinds of leaves or *phytoelements* (twigs, fruits, etc.), define 'leaf' area densities $\eta_p(\Omega_L, \mathbf{r})$, and optical coefficients $\sigma_p$ and $\varepsilon_p$, for each kind $p$ of phytoelement. Then the canopy optical coefficients involve sums over $p$.

(ii) To allow the frequency $\nu$ and polarization $\pi$ of photons to get changed by scatterings off leaves, let $\Omega$ include $\nu$ and $\pi$, besides directions.

(iii) Note finally that horizontally non-homogeneous situations can also be treated by letting $\Omega$ include horizontal coordinates (e.g., points of entry into the canopy at height $z_0$).

**Appendix A**: **Functions of a $2\times 2$ matrix**

Denote, for any $2\times 2$ matrix $\mathbf{K}$, and any analytic function $f(x)$:

$$k = \tfrac{1}{2}Tr\mathbf{K}, \qquad \overline{\mathbf{K}} = \mathbf{K} - k\mathbf{1}, \qquad \lambda = \sqrt{-\det \overline{\mathbf{K}}}$$
$$f(\mathbf{K}) = f(k + \overline{\mathbf{K}}) = \sum_{n=0}^{\infty} \tfrac{1}{n!} f^{(n)}(k)\overline{\mathbf{K}}^n, \qquad f^{(n)}(x) \equiv d^n f/dx^n \tag{A.1}$$

We now have, since $Tr\overline{\mathbf{K}} = 0$:

$$\overline{\mathbf{K}} = \begin{pmatrix} a & b \\ c & -a \end{pmatrix}, \qquad \overline{\mathbf{K}}^2 = \begin{pmatrix} a^2 + bc & 0 \\ 0 & a^2 + bc \end{pmatrix} = -\mathbf{1}\det\overline{\mathbf{K}} = \lambda^2 \mathbf{1} \tag{A.2}$$

so that $\overline{\mathbf{K}}^{2m} = \mathbf{1}\lambda^{2m}$, $\overline{\mathbf{K}}^{2m+1} = \overline{\mathbf{K}}\lambda^{2m}$. It follows that

$$f(\mathbf{K}) = \mathbf{1}\sum_{m=0}^{\infty} \tfrac{1}{(2m)!} f^{(2m)}(k)\lambda^{2m} + \overline{\mathbf{K}}\sum_{m=0}^{\infty} \tfrac{1}{(2m+1)!} f^{(2m+1)}(k)\lambda^{2m}$$
$$= \tfrac{1}{2}[f(k+\lambda) + f(k-\lambda)]\mathbf{1} + \tfrac{1}{2\lambda}[f(k+\lambda) - f(k-\lambda)]\overline{\mathbf{K}} \tag{A.3}$$

In particular:

$$e^{\mathbf{K}} = e^{\tfrac{1}{2}Tr\mathbf{K}}e^{\overline{\mathbf{K}}}, \qquad e^{\overline{\mathbf{K}}} = \mathbf{1}\cosh\lambda + \overline{\mathbf{K}}(\sinh\lambda)/\lambda \tag{A.4}$$

**Appendix B: Some integrals involving $\Omega \cdot \Omega'$**

In this Appendix, we evaluate some integrals involving scalar products

$$\Omega \cdot \Omega' = a + b\cos(\varphi - \varphi'), \qquad a \equiv \mu\mu', \qquad b \equiv \sin\theta\sin\theta' \geq 0 \tag{B.1}$$

where $\sin\theta \geq 0$ since $\theta \in (0,\pi)$. Also, $\mu = \cos\theta \in (-1,1)$, so that

$$\cos^{-1}\mu \in (0,\pi), \qquad \cos^{-1}(-\mu) = \pi - \cos^{-1}\mu \tag{B.2}$$

We will use the function

$$\zeta(\mu) = \sqrt{1-\mu^2} - \mu\cos^{-1}\mu = -\int d\mu \cos^{-1}\mu, \qquad \zeta(-\mu) = \zeta(\mu) + \pi\mu$$
$$\zeta(\cos\theta) = \sin\theta - \theta\cos\theta = \int d\theta\,\theta\sin\theta, \qquad \theta \in (0,\pi), \quad \mu \in (-1,1) \tag{B.3}$$

**B.1** *The functions* $g(\mu,\mu')$: Denote

$$g_\pm(\mu,\mu') = \tfrac{1}{2\pi}\int_{-\pi}^{\pi} d\varphi\, |\Omega \cdot \Omega'|_\pm = \tfrac{1}{2}g(\mu,\mu') \pm \tfrac{1}{2}\mu\mu' \qquad (a)$$
$$g(\mu,\mu') = g_+ + g_- = \tfrac{1}{2\pi}\int_{-\pi}^{\pi} d\varphi\, |\Omega \cdot \Omega'| = \tfrac{1}{2\pi}\int_{-\pi}^{\pi} d\varphi\, |a + b\cos\varphi| \qquad (b) \tag{B.4}$$
$$\Delta g(\mu,\mu') \equiv g_+ - g_- = \tfrac{1}{2\pi}\int_{-\pi}^{\pi} d\varphi\, \Omega \cdot \Omega' = \mu\mu' \qquad (c)$$

Note the integral and special values (since, e.g., $b = 0$ if $\mu' = 1$):

$$\int_{-1}^{1} d\mu\, g(\mu,\mu') = \tfrac{1}{2\pi} \int d\Omega |\mathbf{\Omega} \cdot \mathbf{\Omega}'| = 1 \tag{B.5}$$

$$g(\mu,1) = \tfrac{1}{2\pi}\int_{-\pi}^{\pi} d\varphi\, |\mu| = \hat{\mu}, \qquad g(\mu,0) = \tfrac{1}{2\pi}\int_{-\pi}^{\pi} d\varphi\, |\sin\theta\cos\varphi| = \tfrac{1}{\pi}\sin\theta \tag{B.6}$$

Note also the symmetries, since $a + b\cos\varphi \to -(a + b\cos(\varphi + \pi))$ under $(\mu,\mu') \to (\mu,-\mu')$:

$$g(\mu,\mu') = g(\mu',\mu) = g(-\mu,\mu') = g(\mu,-\mu') = g(-\mu,-\mu') \tag{B.7}$$

If $\hat{a} \geq b$, then $|a + b\cos\varphi| = (\mathrm{sgn}\,a)(a + b\cos\varphi)$ for all $\varphi$, so that $g(\mu,\mu') = (\mathrm{sgn}\,a)a = \hat{a}$. If $\hat{a} \leq b$, then, since $b \geq 0$, and $b\cos\varphi + a < 0$ for $\cos\varphi < -a/b$:

$$g_+(\mu,\mu') = \tfrac{1}{2\pi}\int_{-\varphi_c}^{\varphi_c} d\varphi (b\cos\varphi + a) = \tfrac{1}{\pi}(b\sin\varphi_c + a\varphi_c), \qquad \varphi_c \equiv \cos^{-1}(-a/b) \tag{B.8}$$

Hence, noting that $\sin\varphi_c = \sqrt{1 - (a/b)^2}$, and using (B.3):

$$g(\mu,\mu') = 2g_+ - \Delta g = \begin{cases} \hat{\mu}\hat{\mu}' & \text{if } \hat{a} \geq b \\ \tfrac{2}{\pi} b\zeta(-a/b) - a & \text{if } \hat{a} \leq b \end{cases} \tag{B.9}$$

Note that the symmetries (B.7) follow from $\zeta(-\mu) = \zeta(\mu) + \pi\mu$.

**B.2 The functions $g_{uu}$:** We next evaluate ($u$ stands for $+$ or $-$)

$$\begin{aligned}
g_{u_i,u_f}(\mu_i,\mu_f,\mu_L,\varphi_{if}) &\equiv \int_{-\pi}^{\pi} d\varphi_L |\mathbf{\Omega}_i \cdot \mathbf{\Omega}_L|_{u_i} |\mathbf{\Omega}_f \cdot \mathbf{\Omega}_L|_{u_f} \\
&= b_i b_f \int_{-\pi}^{\pi} d\varphi_L |\alpha_i + \cos\varphi_L|_{u_i} |\alpha_f + \cos(\varphi_L + \varphi_{if})|_{u_f}
\end{aligned} \tag{B.10}$$

where $\varphi_{if} \equiv \varphi_i - \varphi_f$, and $\alpha_i = a_i/b_i$, $\alpha_f = a_f/b_f$, with

$$a_i = \mu_i\mu_L, \quad b_i = \sin\theta_i\sin\theta_L, \quad a_f = \mu_f\mu_L, \quad b_f = \sin\theta_f\sin\theta_L \tag{B.11}$$

The integration (B.10) is over (one or two) intervals $(\varphi_{L1},\varphi_{L2})$ wherein $u_i(\alpha_i + \cos\varphi_L) > 0$ and $u_f(\alpha_f + \cos(\varphi_L + \varphi_{if})) > 0$ simultaneously. Hence, $b_i^{-1}b_f^{-1}g_{u_i,u_f}$ is a sum of (one or two) terms $f_{\alpha_i}^{\alpha_f}(\varphi,\varphi_{L2}) - f_{\alpha_i}^{\alpha_f}(\varphi,\varphi_{L1})$, where $f_{\alpha_i}^{\alpha_f}(\varphi,\varphi_L)$ is the indefinite integral

$$\begin{aligned}
f_{\alpha_i}^{\alpha_f}(\varphi,\varphi_L) &= \int d\varphi_L (\alpha_i + \cos\varphi_L)(\alpha_f + \cos(\varphi_L - \varphi)) \\
&= (\alpha_i\alpha_f + \tfrac{1}{2}\cos\varphi)\varphi_L + \alpha_i\sin(\varphi_L - \varphi) + \alpha_f\sin\varphi_L + \tfrac{1}{4}\sin(2\varphi_L - \varphi)
\end{aligned} \tag{B.12}$$

Thus $g_{u_i,u_f}$ is not simple, but is known analytically, and easy to program on the computer.

**B.3** *Other integrals*: We now show that

$$\int d\Omega_L \left|\Omega_i \cdot \Omega_L\right|_{u_i} \left|\Omega_f \cdot \Omega_L\right|_{u_f} = \tfrac{2}{3}\zeta\!\left(-u_i u_f \Omega_i \cdot \Omega_f\right) \tag{B.13}$$

$$\int_{-\pi}^{\pi} d\varphi_L \left|\Omega_i \cdot \Omega_L\right|_{u_i} \left|\Omega_f \cdot \Omega_L\right|_{u_f} = \tfrac{1}{2}\sin\theta_i \sin\theta_f\, \zeta\!\left(-u_i u_f \cos(\varphi_f - \varphi_i)\right) \quad \text{if} \quad \mu_L = 0 \tag{B.14}$$

We first show that

$$\chi_{u_i,u_f}(\phi) \equiv \int_{-\pi}^{\pi} d\varphi_L \left|\cos\varphi_L\right|_{u_i} \left|\cos(\varphi_L - \phi)\right|_{u_f} = \tfrac{1}{2}\zeta(-u_i u_f \cos\phi) \tag{B.15}$$

Since $\cos(-\varphi) = \cos\varphi$ and $\cos(\varphi + \pi) = \cos(\pi - \varphi) = -\cos\varphi$, we have

$$\chi_{u_i,u_f}(\phi) = \chi_{u_i,u_f}(-\phi), \qquad \chi_{-u_i,u_f}(\phi) = \chi_{u_i,-u_f}(\phi) = \chi_{u_i,u_f}(\pi - \phi) \tag{B.16}$$

Hence it suffices to evaluate $\chi_{-+}(\phi)$ for $\phi \in (0,\pi)$. Putting $\varphi_L - \tfrac{1}{2}\phi = \overline{\varphi}_L$, using $\cos x \cos y = \tfrac{1}{2}\cos(x+y) + \tfrac{1}{2}\cos(x-y)$, and (B.3), we get, for $\phi \in (0,\pi)$:

$$\begin{aligned}
\chi_{-+}(\phi) &= \int d\overline{\varphi}_L \left|\cos(\overline{\varphi}_L + \tfrac{1}{2}\phi)\right|_{-} \left|\cos(\overline{\varphi}_L - \tfrac{1}{2}\phi)\right|_{+} \\
&= -\tfrac{1}{2}\int_{\frac{1}{2}\pi-\frac{1}{2}\phi}^{\frac{1}{2}\pi+\frac{1}{2}\phi} d\overline{\varphi}_L \left[\cos 2\overline{\varphi}_L + \cos\phi\right] = \tfrac{1}{2}(\sin\phi - \phi\cos\phi) = \tfrac{1}{2}\zeta(\cos\phi)
\end{aligned} \tag{B.17}$$

We used $\cos(\overline{\varphi}_L + \tfrac{1}{2}\phi) < 0 \Leftrightarrow \overline{\varphi}_L \in (\tfrac{\pi}{2} - \tfrac{\phi}{2}, \tfrac{3\pi}{2} - \tfrac{\phi}{2})$ and $\cos(\overline{\varphi}_L - \tfrac{1}{2}\phi) > 0 \Leftrightarrow \overline{\varphi}_L \in (-\tfrac{\pi}{2} + \tfrac{\phi}{2}, \tfrac{\pi}{2} + \tfrac{\phi}{2})$, the intersection of the two intervals being $(\tfrac{\pi}{2} - \tfrac{\phi}{2}, \tfrac{\pi}{2} + \tfrac{\phi}{2})$, if $\phi \in (0,\pi)$.

To prove (B.13), note that its left side is a function of the angle between $\Omega_i$ and $\Omega_f$, so that we may choose coordinate axes such that $\Omega_i = (\mu_i = 0, \varphi_i = 0)$, $\Omega_f = (\mu_f = 0, \varphi_f = \phi)$ where $\cos\phi = \Omega_i \cdot \Omega_f$. Then $\Omega_i \cdot \Omega_L = \sqrt{1 - \mu_L^2}\cos\varphi_L$, $\Omega_f \cdot \Omega_L = \sqrt{1 - \mu_L^2}\cos(\varphi_L - \phi)$, so that

$$(\text{B.13}) = \int_{-1}^{1} d\mu_L (1 - \mu_L^2) \int_{-\pi}^{\pi} d\varphi_L \left|\cos\varphi_L\right|_{u_i} \left|\cos(\varphi_L - \phi)\right|_{u_f} = \tfrac{4}{3}\chi_{u_i,u_f}(\phi) \tag{B.18}$$

As to (B.14), it follows from $\Omega \cdot \Omega_L = \sin\theta\cos(\varphi_L - \varphi)$ if $\mu_L = 0$, and (B.15).

We finally note the indefinite integrals ($\mu \equiv \cos\theta$):

$$\begin{aligned}
f_1(\theta) &\equiv \int d\mu\, \zeta(\mu) = \int (\sin\theta) d\theta (\sin\theta - \theta\cos\theta) = \tfrac{1}{2}\theta - \tfrac{3}{8}\sin 2\theta + \tfrac{1}{4}\theta\cos 2\theta \\
f_2(\theta) &\equiv \int d\theta\, \zeta(\cos\theta) = -2\cos\theta - \theta\sin\theta, \qquad \theta \in (0,\pi)
\end{aligned} \tag{B.19}$$

whence the definite integrals

$$\int_{-1}^{1} d\mu\, \zeta(\pm\mu) = f_1(\pi) - f_1(0) = \tfrac{3}{4}\pi, \qquad \int d\Omega\, \zeta(\pm\Omega\cdot\Omega') = \tfrac{3}{2}\pi^2$$
$$\int_{-\pi}^{\pi} d\varphi\, \zeta(\cos\varphi) = 2\int_{0}^{\pi} d\theta\, \zeta(\cos\theta) = 2f(\pi) - 2f(0) = 8 \tag{B.20}$$

By using (B.20) and $\int_{0}^{\pi} d\theta_f \sin^2\theta_f = \tfrac{1}{2}\pi$, one can verify (82) directly on (71) and (70).